\documentstyle[epsf]{aipproc}

\def\lattbootplottwo#1#2{\centering \leavevmode
    \epsfxsize=.48\hsize \epsfbox{#1} \hfil
    \epsfxsize=.48\hsize \epsfbox{#2}}

\def\lattbootplotfiddle#1#2#3#4#5#6#7{\centering \leavevmode
    \vbox to#2{\rule{0pt}{#2}}
    \includegraphics{#1}}

\begin{document}

\title{Nucleosynthesis of Elements in Low to Intermediate Mass 
 Stars through the AGB Phase}

\author{John C. Lattanzio and Arnold I. Boothroyd}

\address{Dept.\ of Mathematics, Monash University, Clayton, VIC.\ 3168,
 Australia}

\maketitle

\begin{abstract}

We present a review of the main phases of stellar evolution
with particular emphasis on the nucleosynthesis and mixing mechanisms in
low- and intermediate-mass stars. 
In addition to explicit studies of the
effects of the first, second and third dredge-up, we also 
discuss {\it cool bottom processing\/} and {\it hot bottom burning}. 

\end{abstract}

\section{Introduction}  \label{S:lattboot:intro}

In recent years a wealth of new abundance data has been obtained,
both from stellar abundance observations and from precise laboratory
measurements of isotope ratios in stellar grains from meteorites.  This
places strong constraints on nucleosynthesis and mixing in low and
intermediate mass stars.  A quantitative understanding relies on our
knowledge of both the stellar evolution and the nucleosynthesis
occurring in the stars.

There are specific phases of a star's life where mixing brings to the
surface the products of interior nucleosynthesis.  These are referred to
as ``dredge-up'' events, and there are basically three, although the
details are mass-dependent.
There are also observations which require mixing beyond what is
found in the standard theory.  This paper aims to review
these events, both qualitatively, for non-experts,
and quantitatively, for those more familiar with the area.

\section{Basic Stellar Evolution} \label{S:lattboot:evolution}

In this section we give a qualitative overview of the evolution of stars
of masses 1 and $5\> M_\odot$, with emphasis on the
mechanisms and phenomenology of the structural and evolutionary changes.
We consider in detail the evolution up to the beginning of thermal
pulses on the AGB (the ``TP-AGB''), with a brief discussion of
further evolution (which is discussed in detail in
section~\ref{S:lattboot:TPAGB}).

\subsection{Basic Evolution at 1 Solar Mass}  \label{S:lattboot:m1evoln}

\begin{figure}
\lattbootplotfiddle{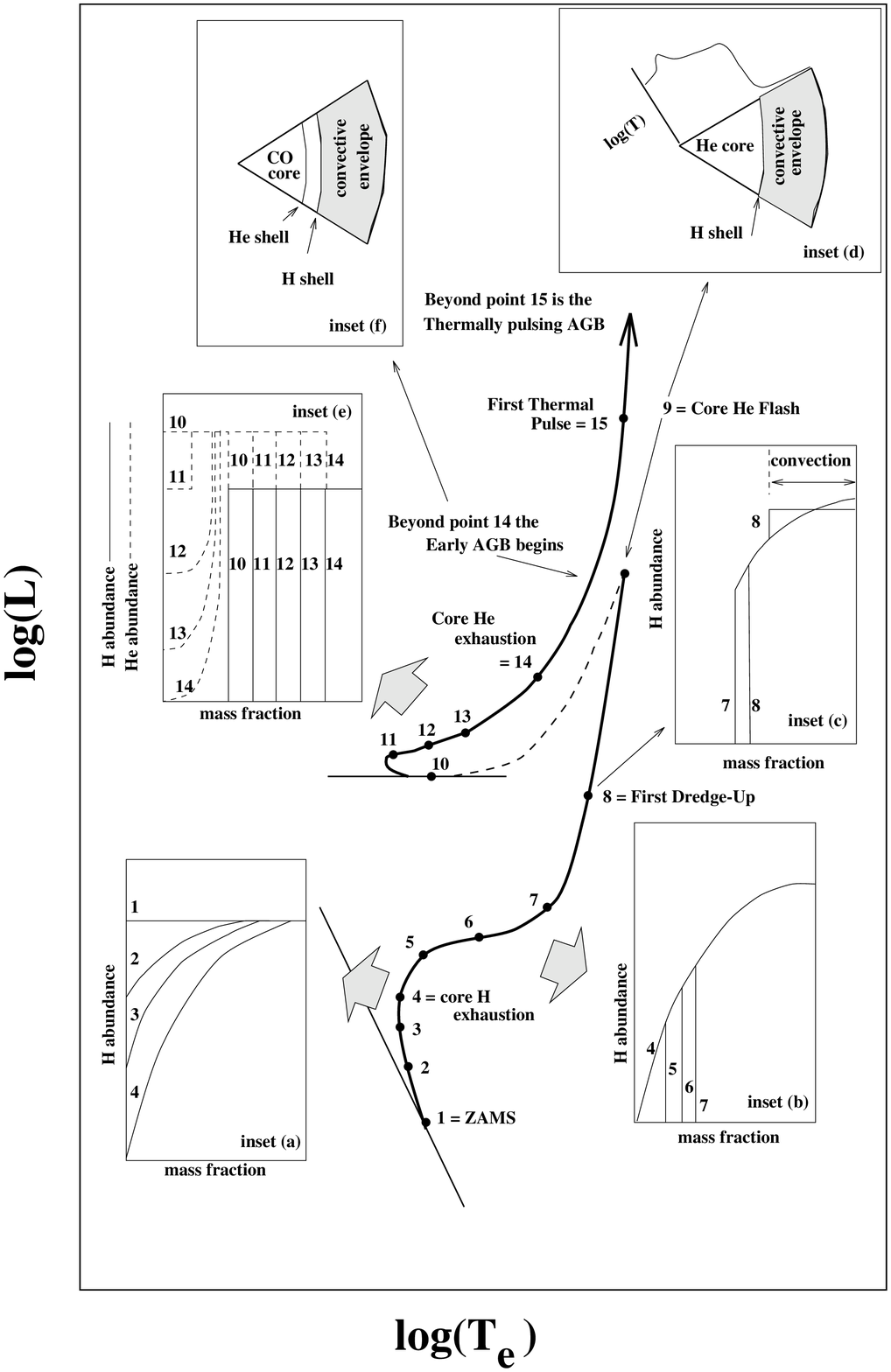}{9 true in}{0}{65}{65}{-190}{70}
\vskip -0.95 true in
\caption{Schematic of evolution at $\sim 1M_\odot$.}
\label{F:lattboot:m1evoln}
\end{figure}

We make the usual assumption that a star reaches the zero-age main
sequence with a homogeneous chemical composition.
Figure~\ref{F:lattboot:m1evoln} shows a schematic HR diagram for a
$\sim 1\>M_\odot$ star.  Core H-burning occurs radiatively, and the
central temperature and density grow in response to the increasing
molecular weight (points 1--3) until central H exhaustion (point~4). The H
profiles are shown in inset~(a) in Figure~\ref{F:lattboot:m1evoln}.
The star now leaves the main sequence and crosses the Hertzsprung Gap
(points 5--7), while the central $^4$He-core becomes electron
degenerate and the nuclear burning is established in a surrounding shell.
Inset~(b) shows the advance of the H-shell during this
evolution.  Simultaneously, the star is expanding and the outer layers
become convective.  As the star reaches the Hayashi limit ($\sim$ point
7), convection extends quite deeply inward (in mass) from the surface,
and the star ascends the (first) red giant branch (RGB)\hbox{}.  The convective
envelope penetrates into the region where partial H-burning has occurred
earlier in the evolution, as shown in inset~(c) of
Figure~\ref{F:lattboot:m1evoln}.  This material is still mostly~H, but
with added $^4$He~together with the products of CN cycling, primarily
$^{14}$N and~$^{13}$C\hbox{}.  These are now mixed to the surface (point~8);
this phase is known as the ``{\it first dredge-up\/}''.  The most important
surface abundance changes are an increase in the $^4$He~mass fraction
by~$\sim 0.03$ (for masses $\lesssim 4\>M_\odot$), while
$^{14}$N~increases at the expense of $^{12}$C by~$\sim 30$\%, and the
number ratio $^{12}$C/$^{13}$C~drops from its initial value of~$\sim 90$
to lie between 18 and 26~\cite{lattboot:Char94}.  Further details are
given in section~\ref{S:lattboot:dr12} below.

%

As the star ascends the giant branch the $^4$He-core continues to
contract and heat.  Neutrino energy losses from the centre cause the
temperature maximum to move outward, as shown in inset~(d) of
Figure~\ref{F:lattboot:m1evoln}.  Eventually triple-alpha reactions are
ignited at this point of maximum temperature, but with a degenerate
equation of state.  The temperature and density are decoupled, and the
resulting ignition is violent --- the ``core helium
flash'' (point~9: see, e.g.,~\cite{lattboot:D84}).  Following this,
the star quickly moves to the Horizontal Branch where it burns
$^4$He~gently in a convective core, and H~in a shell (which provides
most of the luminosity).  This corresponds to points 10--13 in
Figure~\ref{F:lattboot:m1evoln}.  Helium burning increases the mass
fraction of $^{12}$C and~$^{16}$O (the latter through
$^{12}$C$(\alpha,\gamma)^{16}$O) and the outer regions of the
convective core become stable to the Schwarzschild convection criterion
but unstable to that of Ledoux:\ a situation referred to as
``semiconvection'' (space prohibits a discussion of this phenomenon, but
an excellent physical description is contained
in~\cite{lattboot:Cetal71a,lattboot:Cetal71b}).  The semiconvection
causes the composition profile to adjust itself to produce convective
neutrality, with the resulting profiles as shown in inset~(e) of
Figure~\ref{F:lattboot:m1evoln}.

Following $^4$He exhaustion (point 14), the star ascends the giant
branch for the second time, and this is known as the Asymptotic Giant
Branch, or AGB, phase.  The final proportions of $^{12}$C~and $^{16}$O
in the $^4$He-exhausted core depend on the uncertain rate for the
$^{12}$C$(\alpha,\gamma)$$^{16}$O reaction.  The core now becomes
electron degenerate, and the star's energy output is provided by the
$^4$He-burning shell (which lies immediately above the C-O core) and the
H-burning shell. Above both is the deep convective envelope.  This
structure is shown in inset~(f) in Figure~\ref{F:lattboot:m1evoln}.  We
will later see that the $^4$He-shell is thermally unstable, as witnessed
by the recurring ``thermal pulses''.  Thus the AGB is divided into two
regions: the early-AGB (E-AGB), prior to (and at lower luminosities
than) the first thermal pulse, and the thermally-pulsing AGB (TP-AGB)
beyond this. We will return to this in
section~\ref{S:lattboot:TPAGB}.

\subsection{Basic Evolution at 5 Solar Masses} \label{S:lattboot:m5evoln}

\begin{figure}
\lattbootplotfiddle{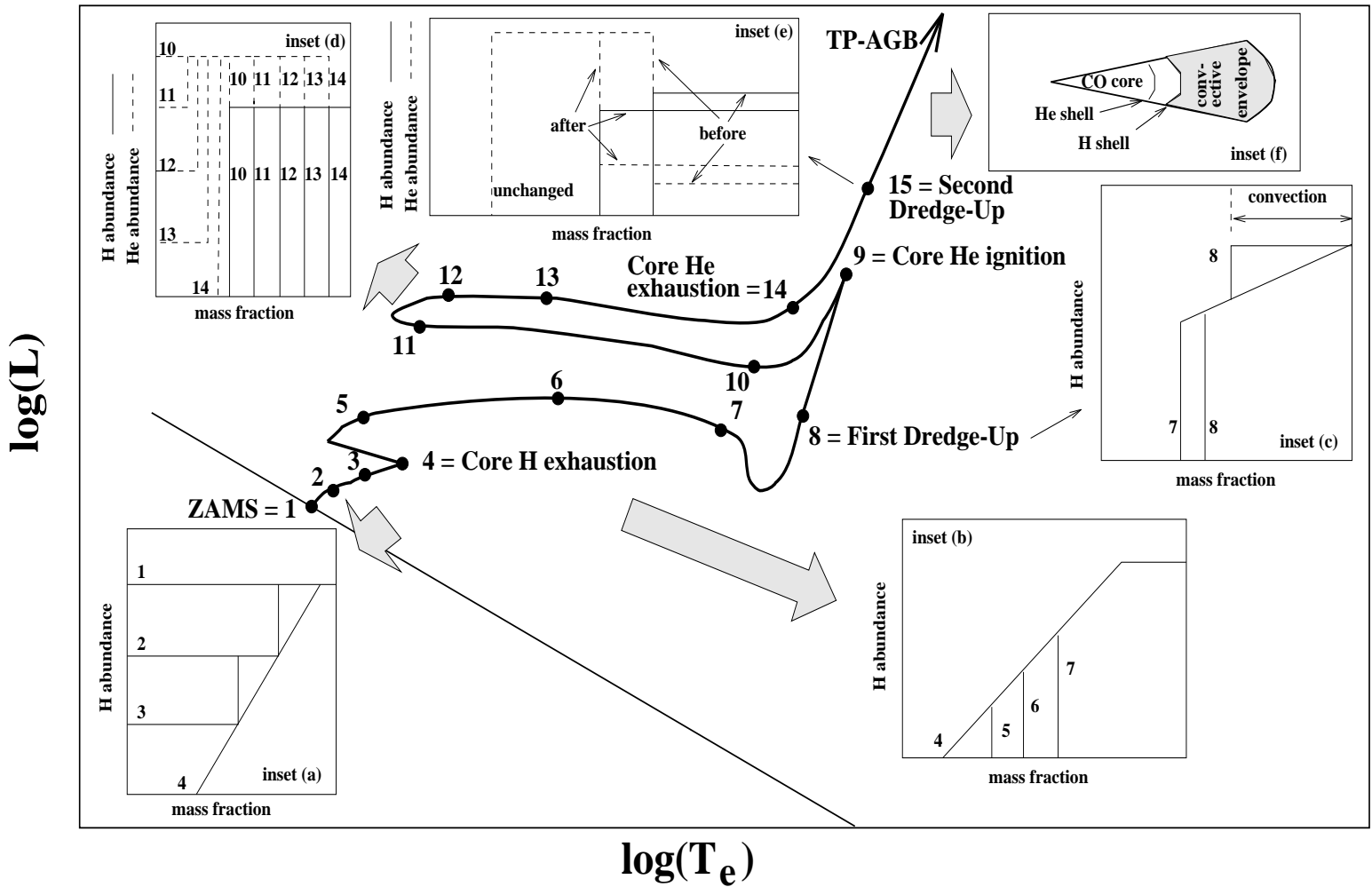}{9 true in}{0}{75}{68}{-180}{80}
\vskip -1.1 true in
\caption{Schematic of evolution at $\sim 5M_\odot$.}
\label{F:lattboot:m5evoln}
\end{figure}

A more massive star, say of $5M_\odot$, begins its life very
similarly to the lower mass star discussed above.  The main initial
difference is that the higher temperature in the core causes CNO cycling
to be the main source of H-burning, and the high temperature dependence
of these reactions causes a convective core to develop.  As H is burned
into $^4$He the opacity (mainly owing to electron scattering, and hence
proportional to the H content) decreases and the extent of the
convective core decreases with time.  This corresponds to points 1--4 in
Figure~\ref{F:lattboot:m5evoln}.  Following core H~exhaustion there is a
phase of shell burning as the star crosses the Hertzsprung Gap (points
5--7 and inset~(b)), and then ascends the (first) giant branch. Again
the inward penetration of the convective envelope (point 8) reaches
regions where there has been partial H-burning earlier in the evolution,
and thus these products (primarily $^{13}$C~and $^{14}$N, produced at
the cost of $^{12}$C) are mixed to the surface in the first dredge-up,
just as seen at lower masses, and sketched in inset~(c) of
Figure~\ref{F:lattboot:m5evoln}.

%

For these more massive stars the ignition of $^4$He~occurs in the centre
and under non-degenerate conditions, and the star settles down to a
period of quiescent $^4$He-burning in a convective core, together with
H-burning in a shell (see inset~(d) in Figure~\ref{F:lattboot:m5evoln}).
The competition between these two energy sources determines the
occurrence and extent of the subsequent blueward excursion in the HR
diagram~\cite{lattboot:L71}, when the star crosses the instability strip
and is observed as a Cepheid variable (points 10--14).  Following core
$^4$He~exhaustion, the structural re-adjustment to shell $^4$He burning
results in a strong expansion, and the H-shell is extinguished as the
star begins its ascent of the AGB\hbox{}.  With this entropy barrier removed,
the inner edge of the convective envelope is free to penetrate the
inactive H-shell; the products of complete H-burning are mixed to
the surface in ``{\it second dredge-up\/}'' (point~15).
This again alters the surface compositions of $^4$He, $^{12}$C, $^{13}$C,
and~$^{14}$N, and reduces the mass of the H-exhausted core,
(the process of mixing $^4$He~outward also mixes H~inward:
see inset~(e) in Fig.~\ref{F:lattboot:m5evoln}).  There is
a critical mass (of $\sim 4\>M_\odot$, but dependent on composition)
below which the second dredge-up does not occur.  Following second
dredge-up, the H-shell is re-ignited, and the first thermal pulse occurs
soon after:\ the star has reached the thermally-pulsing AGB, or
TP-AGB\hbox{}.  Note that at this stage the structure is qualitatively
similar for all masses.

\subsection{The Key Mixing Events}  \label{S:lattboot:mixing}

As we saw above, when a star approaches the RGB after the completion of
main sequence core H-burning, its convective envelope deepens,
eventually dredging up products of partial H-burning ({\it
first dredge-up\/}).  The first dredge-up ceases when the convective
envelope reaches its maximum inward penetration, and then
recedes again.  This leaves behind a sharp composition discontinuity.
For low mass stars ($\lesssim 2.5\>M_\odot$), the H-burning shell
catches up to and erases this discontinuity while the star is still on
the RGB; for higher masses, the star leaves the
RGB before this can take place.  There is observational evidence from
surface abundance changes on the RGB~\cite{lattboot:Gil89,%
lattboot:GilB91,lattboot:HarLS88,lattboot:Lang+86,lattboot:ST92,%
lattboot:Kra+93,lattboot:Kra94,lattboot:Char94}
(see section~\ref{S:lattboot:dr12}) that, once the composition discontinuity
is erased, some form of non-convective ``{\it extra mixing\/}'' takes
place, which transports material from the (relatively cool) bottom of
the convective envelope down close to the H-burning shell (where
nuclear burning can alter its composition) and then up to be mixed back
into the convective envelope.  Boothroyd {\it et al.}~\cite{lattboot:BSW95}
referred to this process as ``{\it cool
bottom processing\/}''; the mixing mechanism is not well
understood, but is frequently assumed to be rotation-induced, e.g.,
meridional circulation~\cite{lattboot:SweM79} and/or shear-induced
turbulence~\cite{lattboot:Char95} (similar to the extra mixing process
on the main sequence that yields large
${}^7$Li-depletions in $\sim 1\>M_\odot$ stars~\cite{lattboot:Pin+89}).
Note that ``extra
mixing'' or ``extra deep mixing'' generally result in ``cool bottom
processing'' and hence in surface abundance changes; these three terms
are used essentially synonymously hereafter.

Low mass stars experience significant mass loss on the RGB (totaling
$\sim 0.2\>M_\odot$), with peak mass loss rates of
$\dot M \sim 10^{-7}\>M_\odot$/yr near the tip of the~RGB; some grain
formation may take place during this stage.

After the completion of core He-burning, helium burns in a shell
surrounding the degenerate carbon--oxygen core; the star ascends the AGB,
and the convective envelope deepens again.  In intermediate mass stars,
the H-burning shell is temporarily extinguished and envelope
convection reaches into and below the position of the
H-shell, bringing more nucleosynthesized material to the surface ({\it
second dredge-up\/}).  This occurs on the early AGB (E-AGB)\hbox{}.
Afterwards, the H-shell re-ignites, and periodic thermal pulses
(or helium shell flashes) occur --- the thermally-pulsing AGB, or TP-AGB
(see section~\ref{S:lattboot:TPAGB}).  The strong nuclear energy
generation in these thermal pulses causes a convective region to grow
outwards from the He-burning shell, mixing the products of partial
He-burning (mostly ${}^{12}$C) and of neutron-capture nucleosynthesis
(``$s$-process isotopes'') outwards almost to the base of the H-burning
shell.  Subsequently, the convective envelope reaches into
the intershell region where the products of He-burning were
deposited and mixes them to the surface ({\it third
dredge-up\/}: see section~\ref{S:lattboot:TDU}).  Note that the third
dredge-up is a repeating phenomenon, occurring after almost every pulse
(except for the first few).  This is in contrast to
the first and second dredge-up events, which occur at most once per star
(low mass stars do not experience the second dredge-up at all).

In stars with masses $\gtrsim 4\>M_\odot$, the convective envelope is
deep enough during the long interpulse periods that it reaches into the
H-burning shell, i.e., nuclear processing takes place at the
bottom of the convective envelope, altering its composition.  This is
known as ``{\it hot bottom burning\/}'' (see section~\ref{S:lattboot:HBB}).

The AGB stage of evolution ends when mass loss has removed
almost all of the star's envelope (the ``planetary nebula'' stage
follows).  In the ``superwind'' which terminates the AGB, mass loss
rates of $\dot M \sim 10^{-4}\>M_\odot$/yr are observed; such
dense outflows from cool stars are favorable sites for grain formation.

\section{The RGB and E-AGB: First and Second Dredge-up, and Extra Mixing}
  \label{S:lattboot:dr12}

\begin{figure}[t]
\lattbootplotfiddle{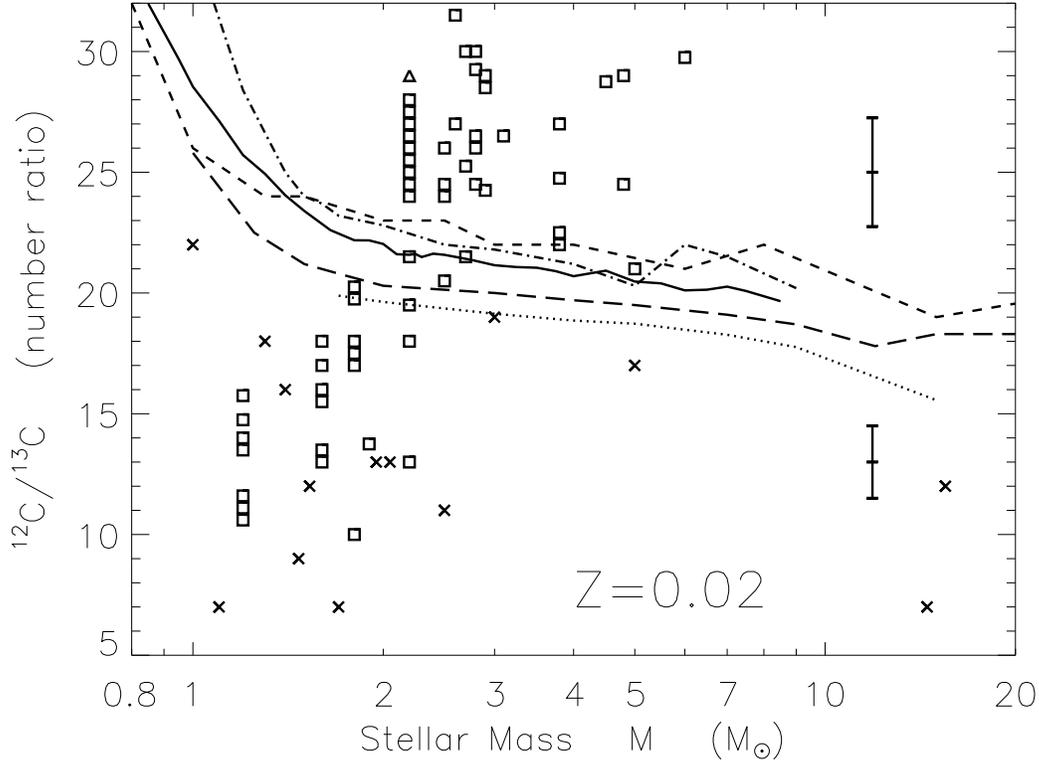}{3.8 true in}{90}{64}{64}{245}{-47}
\caption{Comparison of observed stellar ${}^{12}$C/${}^{13}$C ratios with
theoretical first
dredge-up predictions of various authors (for stars of solar metallicity).
{\it Squares\/}: galactic open cluster
observations~\protect\cite{lattboot:Gil89}
(error-bars at right of plot show typical observational error;
{\it triangle\/} indicates lower limit), having accurate
determinations of the stellar mass.
{\it Crosses\/}: isolated star
observations~\protect\cite{lattboot:HarL84a,lattboot:HarL84b,lattboot:HarLS88},
with masses uncertain by a factor of~$\sim 2$.  Theoretical curves:
{\it solid\/}: Boothroyd \& Sackmann~\protect\cite{lattboot:BS97},
{\it dotted\/}: El~Eid~\protect\cite{lattboot:ElE94},
{\it short-dashed\/}: Dearborn~\protect\cite{lattboot:Dea92},
{\it long-dashed\/}: Schaller {\it et al.}~\protect\cite{lattboot:Schal+92}
and also Charbonnel~\protect\cite{lattboot:Char94},
{\it dot-dashed\/}: Bressan
{\it et al.}~\protect\cite{lattboot:Bre+93}.}
\label{F:lattboot:dr12varcrat}
\end{figure}

Theoretical models of first and second dredge-up without any ``extra mixing''
or ``cool bottom processing''~\cite{lattboot:Dea92,lattboot:Schal+92,%
lattboot:Bre+93,lattboot:Char94,lattboot:ElE94,lattboot:BS97} agree with each
other reasonably well (see, e.g., Fig.~\ref{F:lattboot:dr12varcrat});
results presented here are from the models of Boothroyd \&
Sackmann~\cite{lattboot:BS88,lattboot:SBF90,lattboot:SBK93,lattboot:BS97}.
For solar metallicity ($Z = 0.02$), solar elemental and isotopic abundances
were assumed to represent the initial stellar composition.
The $\alpha$-element
enhancement at lower metallicity was approximated by setting
[O/Fe]${} = -0.5\,$[Fe/H] for [Fe/H]${} > -1$, and constant
[O/Fe]${} = +0.5$ for [Fe/H]${} \le -1$; [C/Fe] and [N/Fe] were taken
to be independent of metallicity, as indicated by observations (see
Timmes {\it et al.}~\cite{lattboot:TimWW95}, and references therein).
The initial isotopic ratios ${}^{12}$C/${}^{13}$C, ${}^{16}$O/${}^{17}$O,
and ${}^{16}$O/${}^{18}$O were taken to be inversely proportional to
Fe/H, as suggested by the galactic chemical evolution models of
Timmes {\it et al.}~\cite{lattboot:TimWW95,lattboot:Tim95} (there are
few observational constraints on the evolution of these isotopic ratios
in the interstellar medium).  Nuclear rates from Caughlan \&
Fowler~\cite{lattboot:CF88} were used, except for $\rm{}^{12}C(\alpha,\gamma)$
(where the rate was multiplied by~1.7, as recommended by Weaver \&
Woosley~\cite{lattboot:WeaW93}), and $\rm{}^{17}O(p,\alpha)$ and
$\rm{}^{17}O(p,\gamma)$, where the 1990 rates of Landr\'e
{\it et al.}~\cite{lattboot:La90} or the (slightly higher) 1995 rates of
Blackmon {\it et al.}~\cite{lattboot:Bla+95,lattboot:Bl96} were used.
A value of the mixing length to pressure scale height ratio of $\alpha = 2.1$
was required in order to obtain a correct model of the
Sun~\cite{lattboot:SBF90,lattboot:SBK93} (note that the value of~$\alpha$
has almost no effect on the depth of dredge-up, as has already been noted by
Charbonnel~\cite{lattboot:Char94}).  Reimers-formula red giant mass
loss~\cite{lattboot:Rei75,lattboot:Kudr78,lattboot:BS97} had negligible
effect, as very little mass had been lost at the time of first
dredge-up (or at the time of second dredge-up, in intermediate mass stars).

Models of ``extra mixing'' and the consequent cool bottom processing on the
RGB~\cite{lattboot:SweM79,lattboot:Char95,lattboot:WBS95,lattboot:DenW96,%
lattboot:BS97,lattboot:SB97}
generally contain free parameter(s) to control the depth and/or speed
of the mixing, whose values may be determined by matching observed RGB
stellar compositions.  Models presented in this 
work~\cite{lattboot:WBS95,lattboot:BS97,lattboot:SB97} use a simple
``conveyor-belt'' circulation scenario, where the extra mixing reaches down
into the outer wing of the H-burning shell.  The temperature
difference $\Delta\log\,T$ between the bottom of mixing and the bottom of
the H-burning shell was considered a free parameter (with a value
$\Delta\log\,T \approx 0.26$ obtained by requiring a $Z = 0.02$,
$1.2\>M_\odot$ case to reproduce the average observed ${}^{12}$C/${}^{13}$C
ratio).  Changes in the envelope structure were followed as the star
climbs the RGB, by using the structure from a stellar evolutionary run
without extra mixing (as the effect of the extra mixing on the envelope
structure should be small).  For carbon and heavier elements, the speed
of circulation is irrelevant within wide limits (a larger number of
rapid circulation passes has the same effect as a smaller number
of slow ones; note that if a diffusive approach to mixing had been
used, as by Denissenkov \& Weiss~\cite{lattboot:DenW96}, then
the choice of 
the speed of mixing would have had a greater effect on the nucleosynthesis).

Charbonnel~\cite{lattboot:Char95} has attempted to go one step beyond
parameterized models such as the one described above, by basing models of
extra RGB mixing on Zahn's~\cite{lattboot:Zahn92} prescription for transport
of chemicals and angular momentum in rotating stars that are losing angular
momentum due to a stellar wind.  Zahn's~\cite{lattboot:Zahn92} approach is
similar to that of Pinsonneault~\cite{lattboot:Pin+89}, assuming a reasonable
functional form for the mixing instabilities and parameterizing them in terms
of a few adjustable parameters.  However, rather than treating all transport
as diffusive, Zahn assumes that meridional circulation can transport both
chemicals and angular momentum, but that shear instabilities lead to largely
horizontal turbulence, homogenizing the star at each radial layer and thus
converting the transport of chemicals to a much slower diffusive process
with diffusion constant $D \approx C_h {3 \over 80\pi} \vert {dJ \over dt}
\vert {1 \over \alpha \rho \Omega r^3}$, where $\alpha = {1 \over 2}
{d\ln\,r^2\Omega \over d\ln\,r}$, $\Omega$~is the local angular velocity,
${dJ \over dt}$~is the rate of loss of angular momentum due to a stellar
wind, and $C_h$~is an adjustable parameter${} \lesssim 1$ --- on the main
sequence,
Zahn~\cite{lattboot:Zahn92} estimates $C_h \approx 0.15$ from the observed
extent of Solar lithium depletion.  Charbonnel~\cite{lattboot:Char95}
assumed $C_h = 1$, with a constant and depth-independent angular rotation
velocity~$\Omega$ corresponding to a reasonable surface rotation velocity
of $1\>$km$\,$sec$^{-1}$ on the RGB, and a Reimers'~\cite{lattboot:Rei75}
mass loss using $\eta = 0.11$ for $Z = 0.001$ and $\eta = 0.035$ for
$Z = 0.0001$ in stars of 0.8 and~$1\>M_\odot$.  The mixing was assumed to
reach down to the ``top of the H-burning shell''~\cite{lattboot:Char95},
presumably determined by the stabilizing effect of the molecular weight
gradient~\cite{lattboot:TalZ97}
in the outer wing of the H-burning shell.  Whether the above
assumptions, and Zahn's parameterization of rotational mixing itself, are
appropriate on the RGB can only be determined by comparison with the
observations, as he noted himself~\cite{lattboot:Zahn92}.

A desirable goal, which has not yet been attempted by any author, would be
a self-consistent model of angular momentum transport,
differential rotation, and the consequent extra mixing throughout a star's
interior during its lifetime (main sequence and RGB, at
the least), using a theoretical prescription of rotational mixing such as
that of Zahn~\cite{lattboot:Zahn92} or of Pinsonneault~\cite{lattboot:Pin+89}.
Complicating such an attempt would be the possibility that the parameter
values --- or even the parameterization --- appropriate to the main sequence
might require changes to correctly describe the RGB\hbox{}.

%
\begin{figure}[t]
\lattbootplottwo{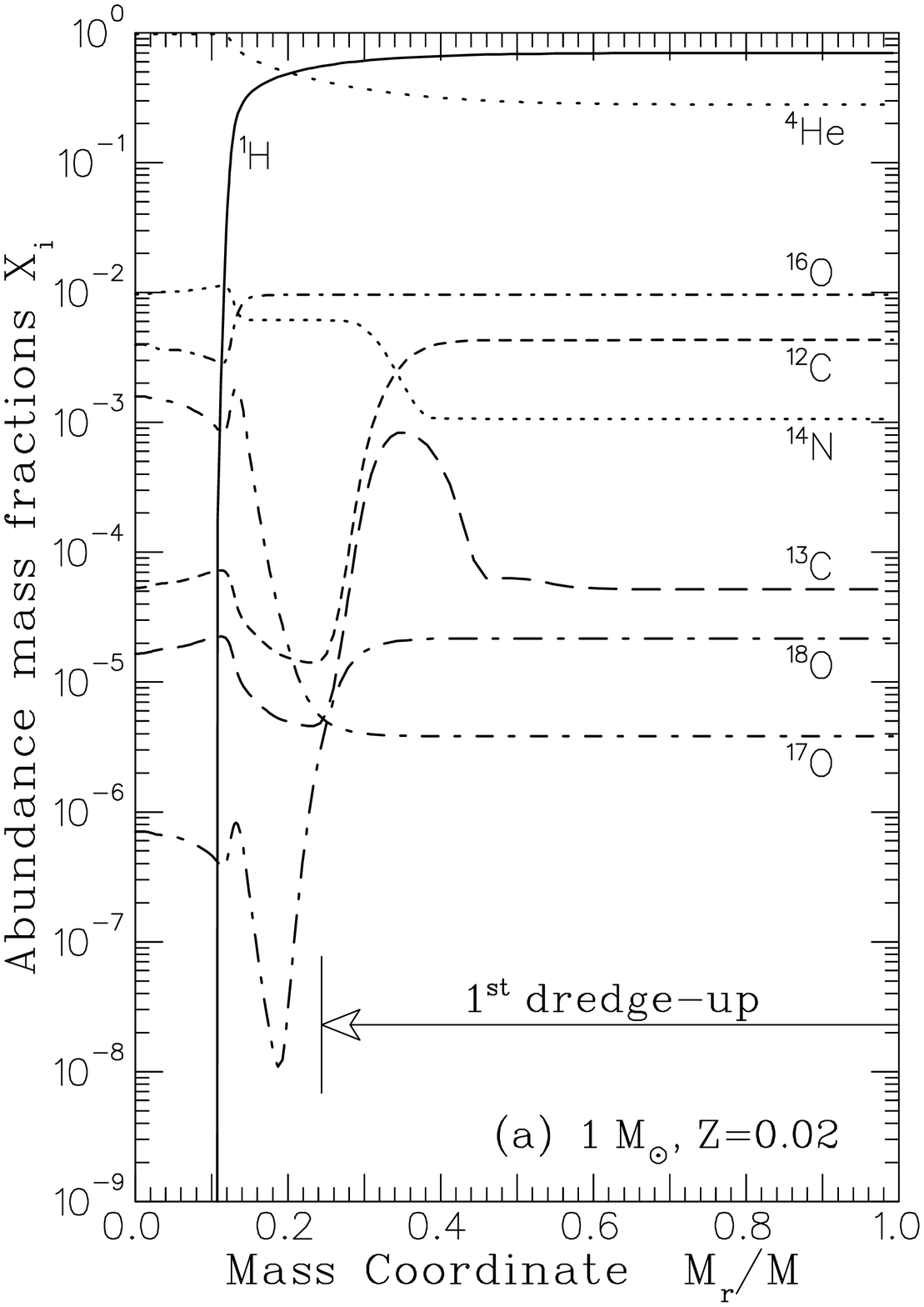}{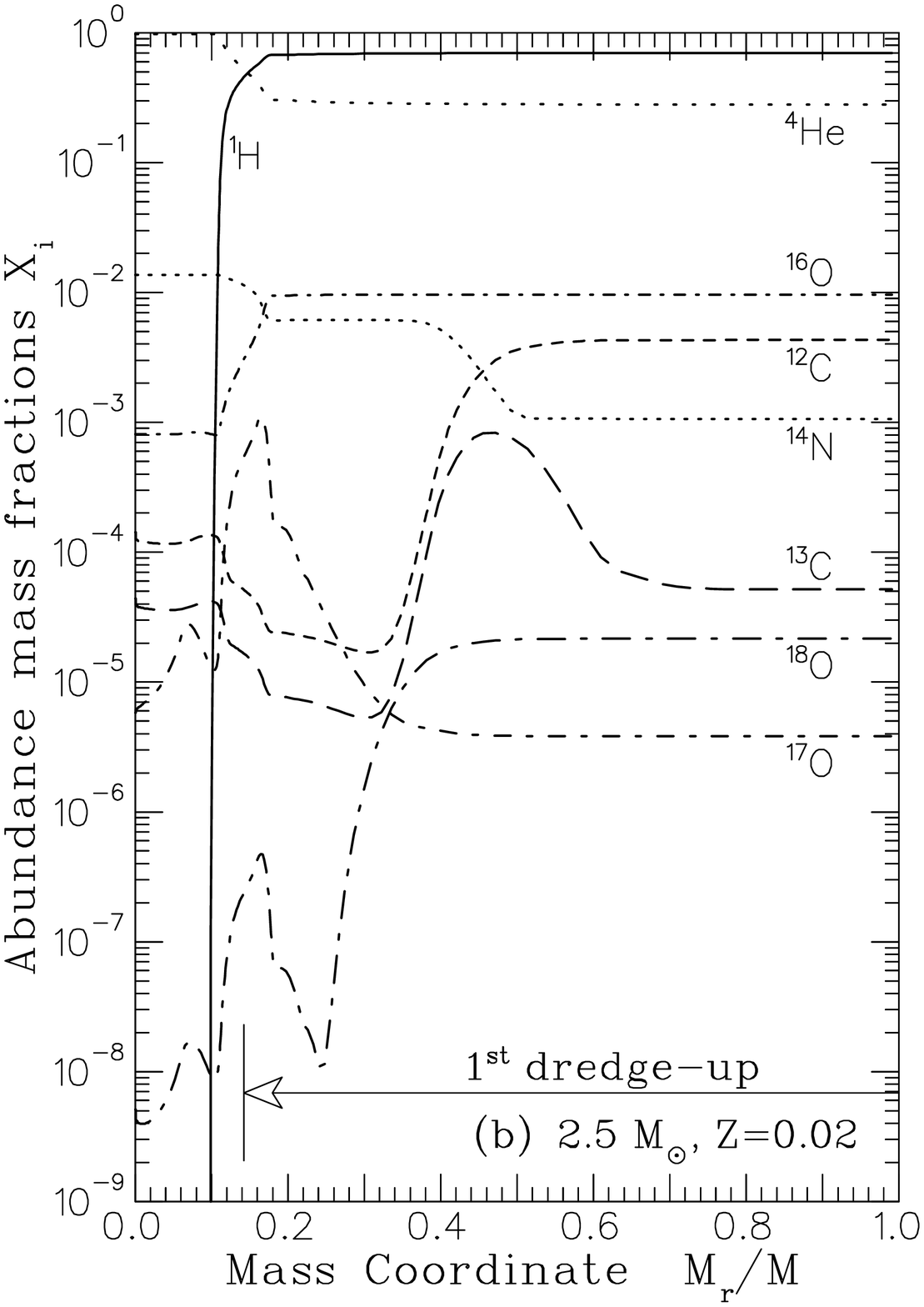}
\caption{Composition profiles as a function of the normalized mass coordinate,
for stars near the base of the RGB (prior to first dredge-up); the depth later
reached by first dredge-up is indicated by the horizontal arrow at the bottom.
({\it a\/})~$1.0\>M_\odot$ star, 
({\it b\/})~$2.5\>M_\odot$ star; both with solar metallicity ($Z = 0.02$).}
\label{F:lattboot:xprof}
\end{figure}

During core H-burning on the main sequence, partial
H-burning in the outer
core produces a region of altered
abundances (see Fig.~\ref{F:lattboot:xprof});
as the star approaches the RGB and
a deep convective envelope develops, this region is engulfed and mixed
into the envelope ({\it first dredge-up\/}).  Li, Be, and~B
have been destroyed in all but the outer layers of the star.  Partial
\hbox{{\it p\/}-{\it p\/}~chain} burning has left a pocket rich
in~${}^3$He.  Slightly further in, most of the ${}^{15}$N has been
destroyed, and a ${}^{13}$C-pocket exists, where
${}^{12}$C/${}^{13}$C approaches its nuclear equilibrium ratio of~$\sim 3$,
but only part of the ${}^{12}$C has been burned.  Below,
most of the ${}^{12}$C and~${}^{13}$C have been converted into~${}^{14}$N,
${}^{18}$O has been destroyed, and~${}^{17}$O~begins to
be enhanced from partial burning of~${}^{16}$O\hbox{}.

%

As the ${}^{13}$C-pocket is engulfed by the convective envelope, the
surface ${}^{12}$C/${}^{13}$C ratio is reduced from its
large initial value ($\sim 90$ for solar compositions) to~$\sim 30$ in 
low mass stars and somewhat
less ($\sim 20$) in intermediate mass stars (see theoretical curves of
Fig.~\ref{F:lattboot:dr12varcrat}).  However, as shown in
Figure~\ref{F:lattboot:dr12varcrat},
observations of RGB and post-RGB ${}^{12}$C/${}^{13}$C ratios in galactic
open clusters~\cite{lattboot:Gil89} indicate the an increasing trend with
stellar mass, not a decreasing one.

Gilroy \& Brown~\cite{lattboot:GilB91} observed ${}^{12}$C/${}^{13}$C ratios
as a function of luminosity on the RGB\hbox{}.  They found that observed and
theoretical ratios agree very well up to and somewhat past the point of
deepest first dredge-up, but that excess ${}^{13}$C began to appear after
the point
where the H-burning shell
reached the composition discontinuity that was left behind by deepest first
dredge-up.  As discussed by Charbonnel~\cite{lattboot:Char94}, this is
consistent with cool bottom processing
due to relatively slow (weak) extra mixing, because the mixing
instability can be stabilized by a molecular weight gradient: the large
molecular weight gradient at the composition discontinuity acts as a
barrier to mixing.
Once the H-burning shell has
reached (and erased) the composition discontinuity, the extra mixing can
transport envelope material down into the outer wing of the H-burning
shell, where H-burning produces a molecular weight gradient
(how deep into the shell the mixing would reach is determined by the
details of the mixing mechanism, but can be estimated by the observed
nucleosynthetic results at the stellar surface).  For stars of
mass $> 2\>M_\odot$, the end of the RGB occurs before the composition
discontinuity has been erased; thus these intermediate-mass stars do
not encounter RGB cool bottom processing.

%
\begin{figure}[t]
\lattbootplotfiddle{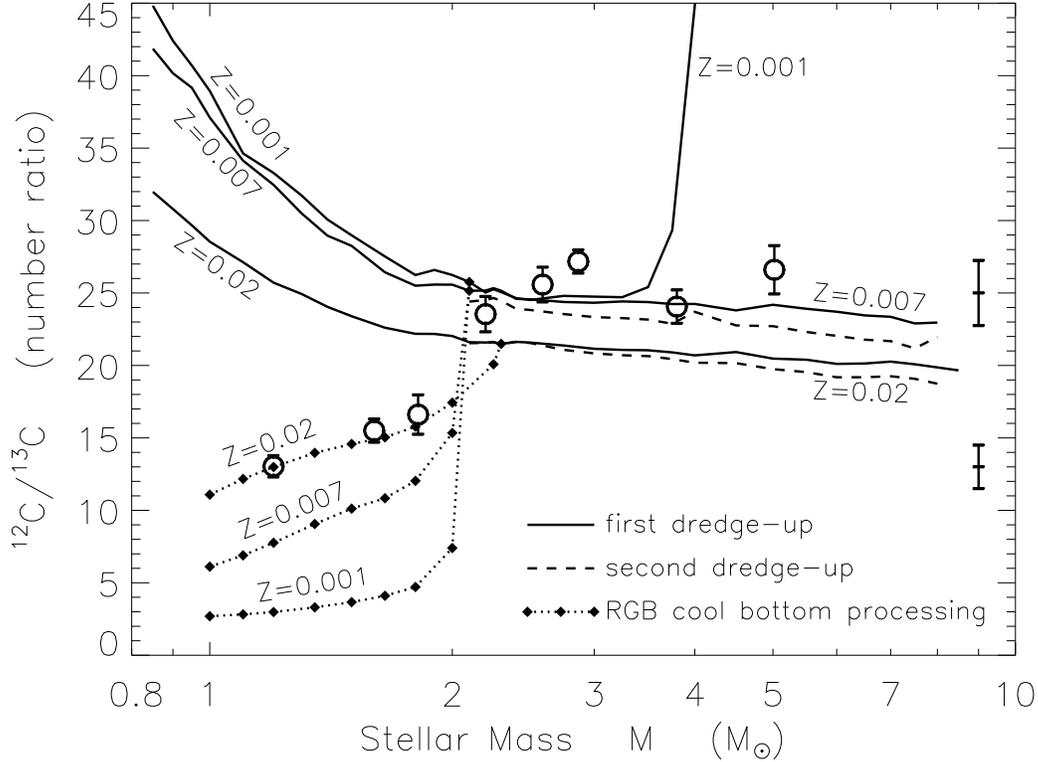}{3.8 true in}{90}{64}{64}{245}{-47}
\caption{Theoretical ${}^{12}$C/${}^{13}$C ratios resulting from first
dredge-up ({\it solid lines\/}), cool bottom processing on the RGB
({\it diamonds\/}), and second dredge-up on the E-AGB in
intermediate-mass stars ({\it dashed lines\/}), for three
metallicities~$Z$ (for clarity, second dredge-up is not plotted for $Z =
0.001$; it coincides approximately with the $Z = 0.007$ case.)  Initial
stellar ${}^{12}$C/${}^{13}$C ratios were assumed to be inversely
proportional to Fe/H, as per galactic chemical evolution models of
Timmes {\it et al.}~\protect\cite{lattboot:Tim95,lattboot:TimWW95}.
{\it Open circles\/} show average ratio from galactic open cluster
observations~\protect\cite{lattboot:Gil89}, with error-bars showing
internal statistical error in the mean (from observational scatter);
error-bars at far right show typical observational errors.  Note cool
bottom processing models were normalized at $1.2\>M_\odot$ (for $Z = 0.02$).}
\label{F:lattboot:dr12crat}
\end{figure}

In Figure~\ref{F:lattboot:dr12crat}, the solid lines show theoretical
predictions of the ${}^{12}$C/${}^{13}$C ratio resulting from first
dredge-up, as a function of stellar mass and metallicity.  The trend
with stellar mass
is due to the fact that low mass stars have narrower ${}^{13}$C-pockets
than intermediate mass stars (the entire ${}^{13}$C-pocket is always
dredged up).  The trend of increased ${}^{12}$C/${}^{13}$C ratio for 
reduced metallicity~$Z$ is due to the fact that the
initial stellar ${}^{12}$C/${}^{13}$C ratio was assumed to be inversely
proportional to Fe/H~\cite{lattboot:Tim95,lattboot:TimWW95}; models where the
initial stellar ${}^{12}$C/${}^{13}$C ratio was assumed to be independent of
metallicity show a very small trend in the opposite direction (see
also Charbonnel~\cite{lattboot:Char94}).
The average observed ${}^{12}$C/${}^{13}$C ratios in RGB and post-RGB stars,
in galactic open clusters of near-solar metallicity, are shown by the open
circles in Figure~\ref{F:lattboot:dr12crat}; for stars of mass
$> 2\>M_\odot$, they are in reasonable agreement with the theoretical curves,
(although they suggest that the ${}^{13}$C-pocket may in fact be about 20\%
smaller than predicted by standard theoretical models).  For stars of mass
$\lesssim 2\>M_\odot$, the observations reflect the ``extra mixing'' and
cool bottom processing that produces additional ${}^{13}$C subsequent to
first dredge-up.  Estimates of the ${}^{12}$C/${}^{13}$C ratio at the tip
of the RGB that result from cool bottom processing, with
the simple circulation model described above,
are shown by the solid diamonds in Figure~\ref{F:lattboot:dr12crat}.
The depth of the extra mixing in the models is determined by the observed
${}^{12}$C/${}^{13}$C ratio in stars of mass $1.2\>M_\odot$, and thus by
definition the cool bottom processing models reproduce that observational
point.  As expected, given the normalization at~$1.2\>M_\odot$,
they also reproduce the trend with stellar mass shown by
the observations of stars with masses between 1.2 and~$2\>M_\odot$.
Under the assumption that extra mixing always reaches the same point in the
outer wing of the H-burning shell, independent of metallicity,
Figure~\ref{F:lattboot:dr12crat} shows
that cool bottom processing has a much greater effect on low
metallicity (Population~II) stars than in stars of near-solar metallicity
(Population~I)\hbox{}.  This is due to the fact that CNO-cycle burning
proceeds at higher temperatures in the H-shell of a
low metallicity star, to compensate for the reduced CNO abundance.  If
anything, these cool bottom processing models underestimate the effect
of metallicity, as discussed below.

The Population~II extra mixing models of Charbonnel~\cite{lattboot:Char95}
reproduce the rapid achievement of $\rm{}^{12}C/{}^{13}C \approx 3$ in the
stellar envelope that is observed in such stars.  A decline in ${}^3$He
(not observable) and in ${}^7$Li (consistent with observations) was also
reported~\cite{lattboot:Char95}.  This behavior is qualitatively correct,
but a quantitative test of the model would require consideration of
Population~I stars, or of more
diagnostic isotopes for such Population~II stars, e.g., total carbon
or oxygen abundances, or heavier elements such as sodium, magnesium, and
aluminum, whose observed behavior on the RGB is discussed below.

Low mass stars ($M \lesssim 2\>M_\odot$), which develop a degenerate
helium core, experience a lengthy RGB stage with significant cool bottom
processing before the violent ignition of core helium, in the
``helium core flash''; stars of higher mass ignite helium quiescently
in a non-degenerate core, and experience only a brief RGB stage, with
no opportunity for cool bottom processing.  In the models of Boothroyd \&
Sackmann~\cite{lattboot:BS97}, Population~II stars of mass
$\gtrsim 4\>M_\odot$ ignite core helium before they ever reach
the RGB, and thus experience no first dredge-up at all (see the $Z = 0.001$
case of Fig.~\ref{F:lattboot:dr12crat}).  Whether this ``early''
core helium ignition actually occurs in higher mass Population~II stars
is not certain: the corresponding models of
Charbonnel~\cite{lattboot:Char94,lattboot:ForC97} show only slightly
shallower first dredge-up than Population~I stars, while the
corresponding models of Lattanzio~\cite{lattboot:Letal97} show intermediate
behavior.  The exact point of core helium
ignition in these stars thus appears to be sensitive to details of the
physical inputs and/or numerical treatment of the stellar models (as are
the form and extent of the ``Cepheid loops'' experienced by
intermediate mass stars during core He-burning).  However, any resulting
uncertainties in the surface composition are wiped out by second dredge-up
on the E-AGB\hbox{}.

As helium begins to burn in a shell surrounding a degenerate carbon--oxygen
core, the star begins to climb the AGB and the convective envelope deepens.
In low mass stars, the convective envelope does not reach as deep as the
H-burning shell, and thus mixes matter that was
already homogenized by first dredge-up; the surface composition thus
changes very little.  In intermediate mass stars, the H-burning
shell is extinguished on the E-AGB and second dredge-up reaches down
into the H-exhausted material left behind by the H-burning
shell.  There almost all the CNO isotopes have been converted
into ${}^{14}$N, so dredge-up of this material has
little effect on the ${}^{12}$C/${}^{13}$C ratio (see
dashed lines in Fig.~\ref{F:lattboot:dr12crat}).  However, if first
dredge-up has not yet mixed the ${}^{13}$C-pocket to the surface (as may be
the case in high mass Population~II stars), second dredge-up will do so.
Since second dredge-up in
intermediate mass stars occurs before any significant amount of mass loss,
the composition of the material injected into the interstellar medium is
not affected by any uncertainty in first dredge-up for
Population~II stars.

The envelope ${}^{14}$N/${}^{15}$N ratio is increased from $\sim 250$ to
$\sim 1000 - 1500$
by first and/or second dredge-up~\cite{lattboot:ElE94,lattboot:ForC97}, due
to the engulfing of ${}^{15}$N-depleted, ${}^{14}$N-enriched material.
However, any cool bottom processing that significantly affects the
${}^{12}$C/${}^{13}$C ratio should also destroy almost all the ${}^{15}$N
in the star's envelope (yielding the nuclear equilibrium ratio
${}^{14}$N/${}^{15}$N${} \gtrsim 10^4$).

%
\begin{figure}[t]
\lattbootplottwo{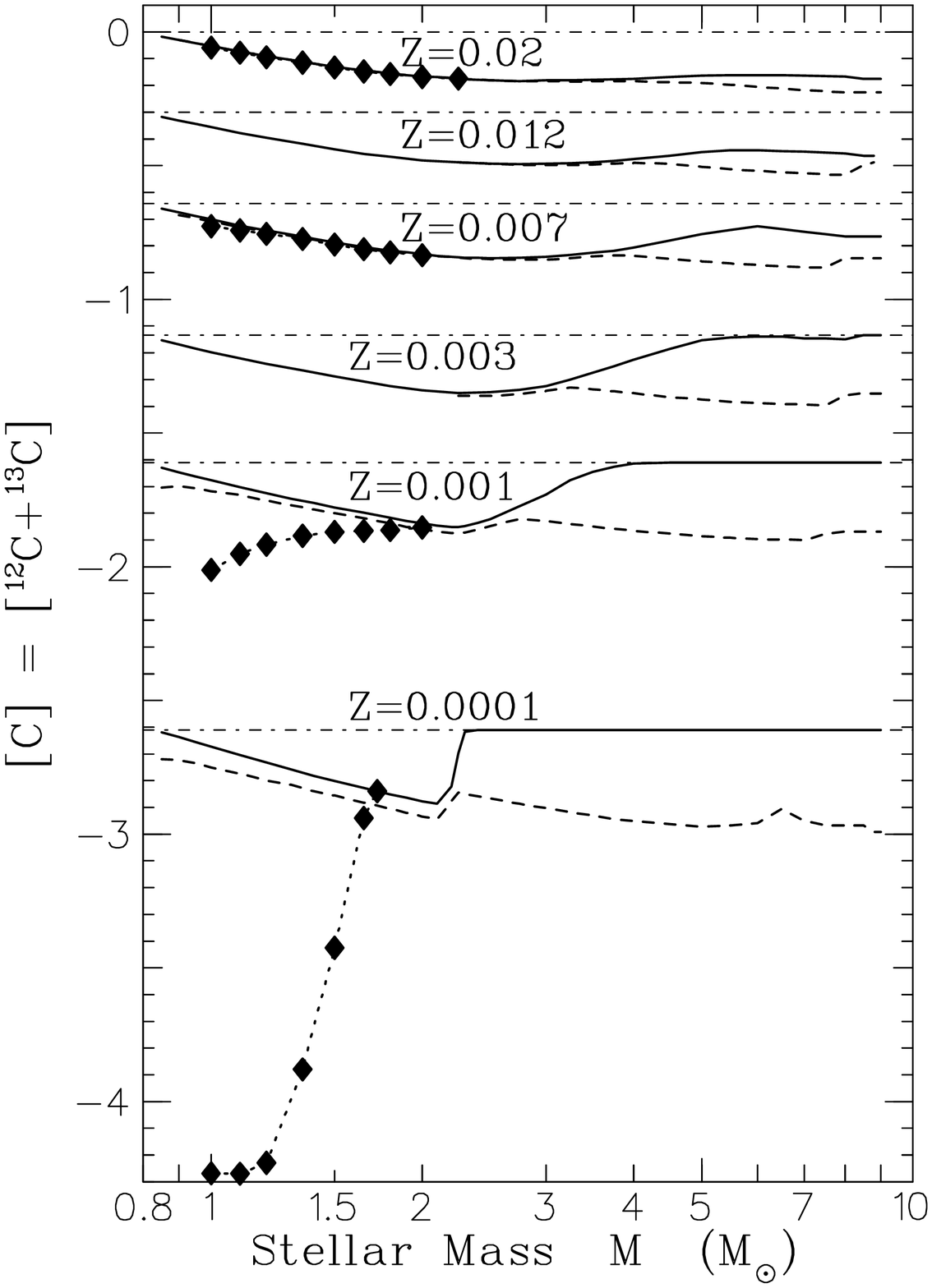}{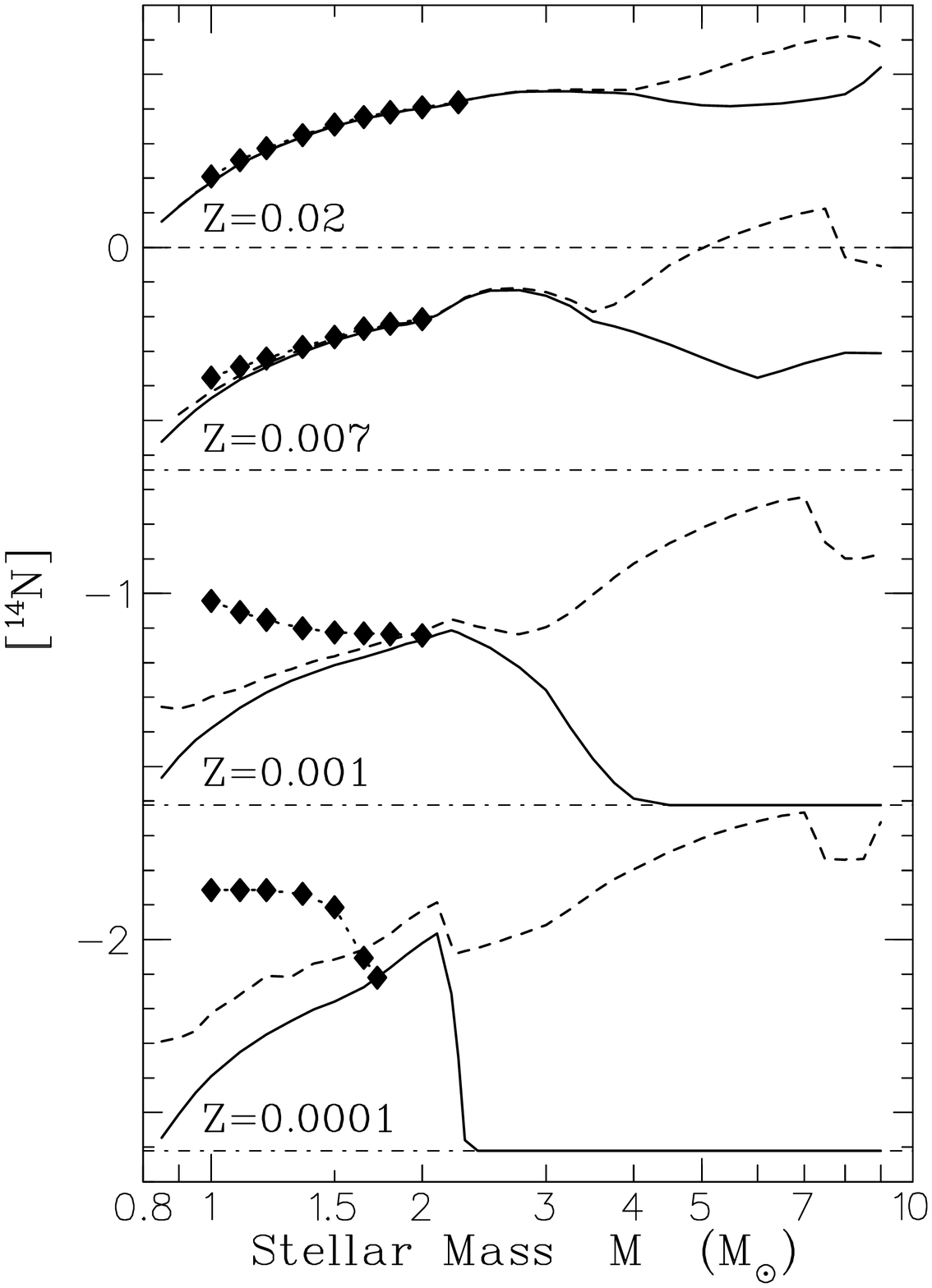}
\caption{Predicted carbon depletions and nitrogen enhancements due to
first dredge-up ({\it solid lines\/}), cool bottom processing on the RGB for
$Z = 0.02$, 0.007, 0.001, and~0.0001 ({\it diamonds\/}), and second dredge-up
on the E-AGB in intermediate-mass stars ({\it dashed lines\/}); note that for
$Z = 0.02$, 0.007, 0.001, and~0.0001, second dredge-up was also computed for
low mass stars, but {\it without\/} including any effects of prior RGB
cool bottom processing.
To avoid confusion, $[{}^{14}{\rm N}]$~curves for $Z = 0.012$ and~0.003 are
omitted.  {\it Dot-dashed lines\/} indicate initial abundances.
Note that $[{\rm C}] \equiv \log\{n({\rm C})/n({\rm C_\odot})\}$, where
number fraction
$n({\rm C}) \equiv n(\hbox{$\rm{}^{12}C$}) + n(\hbox{$\rm{}^{13}C$})$, and that
$[{}^{14}{\rm N}] \equiv \log\{n({}^{14}{\rm N})/n({}^{14}{\rm N}_\odot)\}$.}
\label{F:lattboot:dr12cn}
\end{figure}

Figure~\ref{F:lattboot:dr12cn} illustrates the depletion in total carbon
abundance (by a factor of~$\lesssim 2$) and the corresponding ${}^{14}$N
enhancement due to first and second dredge-up (the total oxygen abundance
is never affected by first or second dredge-up).
Figure~\ref{F:lattboot:dr12cn} also shows that, while the RGB cool bottom
processing models predict negligible effects on total carbon and nitrogen
abundance for near-solar metallicities, large effects are predicted
for Population~II metallicities (these RGB cool bottom processing
models also predict that the total oxygen abundance will not be affected
at any metallicity).  These models agree with observations of Population~II
field stars and of some globular clusters (e.g., M4, 47~Tuc, NGC~3201,
NGC~2298, NGC~288), which show no oxygen depletion (see, e.g.,
Kraft~\cite{lattboot:Kra94},
and references therein).  On the other hand, there are many globular clusters
that do show large oxygen depletions on the RGB (e.g., M5, M13, M3, M92, M15,
M10, NGC~4833, NGC~362: see~\cite{lattboot:Kra94}), and some globular
clusters with metallicities $Z \gtrsim 0.001$ (e.g., 47~Tuc)
exhibit RGB carbon depletions
of over an order of magnitude, much larger than that in the $Z = 0.001$
model in Figure~\ref{F:lattboot:dr12cn}.  A change in the normalization
of extra mixing, i.e., deeper mixing
(down to hotter temperatures), would produce the observed oxygen
depletions (as well as the anti-correlation of O with Na and Al also
observed in some
globular clusters), as shown by the higher-temperature models computed by
Denissenkov \& Weiss~\cite{lattboot:DenW96};
they found, however, that such models could not simultaneously match the carbon
observations (too much carbon was destroyed).
This suggests that there is a star-to-star variation in the depth of extra
mixing, and possibly a variation in the depth of mixing as the star climbs the
RGB\hbox{}.  A similar conclusion follows from
the models of Langer {\it et al.}~\cite{lattboot:LanHS93,lattboot:LanH95}.
Note that the oxygen-depletion and Na-Al observations require that, in some
stars, extra mixing must reach temperatures corresponding to those at the
{\it bottom\/} of the H-burning shell in standard stellar models.
In a case without extra mixing, O depletion and Na production can take place
just outside the burning shell (see, e.g.,~\cite{lattboot:CavSB96}), but when
extra mixing is present higher temperatures are needed to deplete O
significantly over the entire envelope on the RGB
timescale~\cite{lattboot:DenW96}.

%
\begin{figure}[t]
\lattbootplottwo{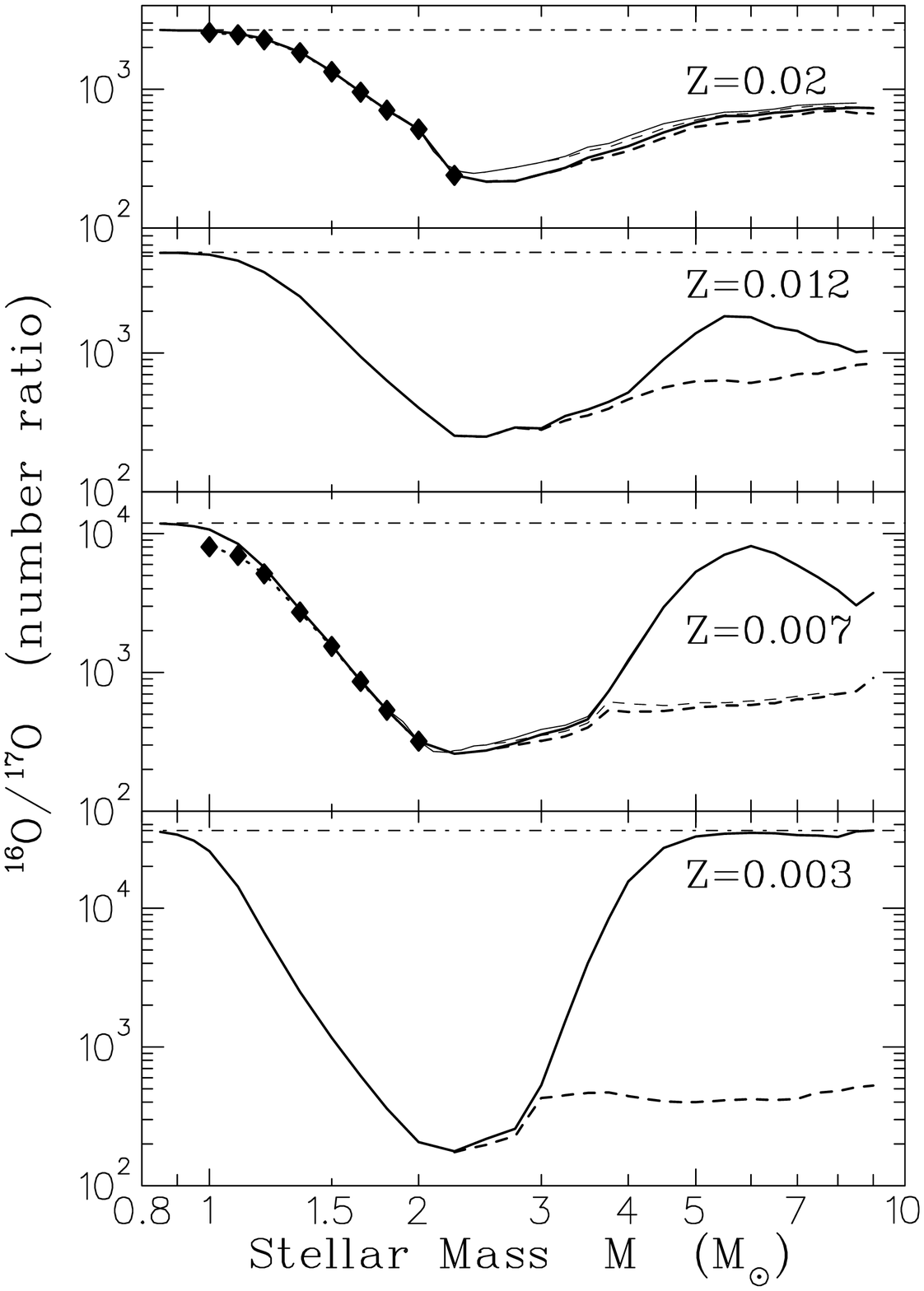}{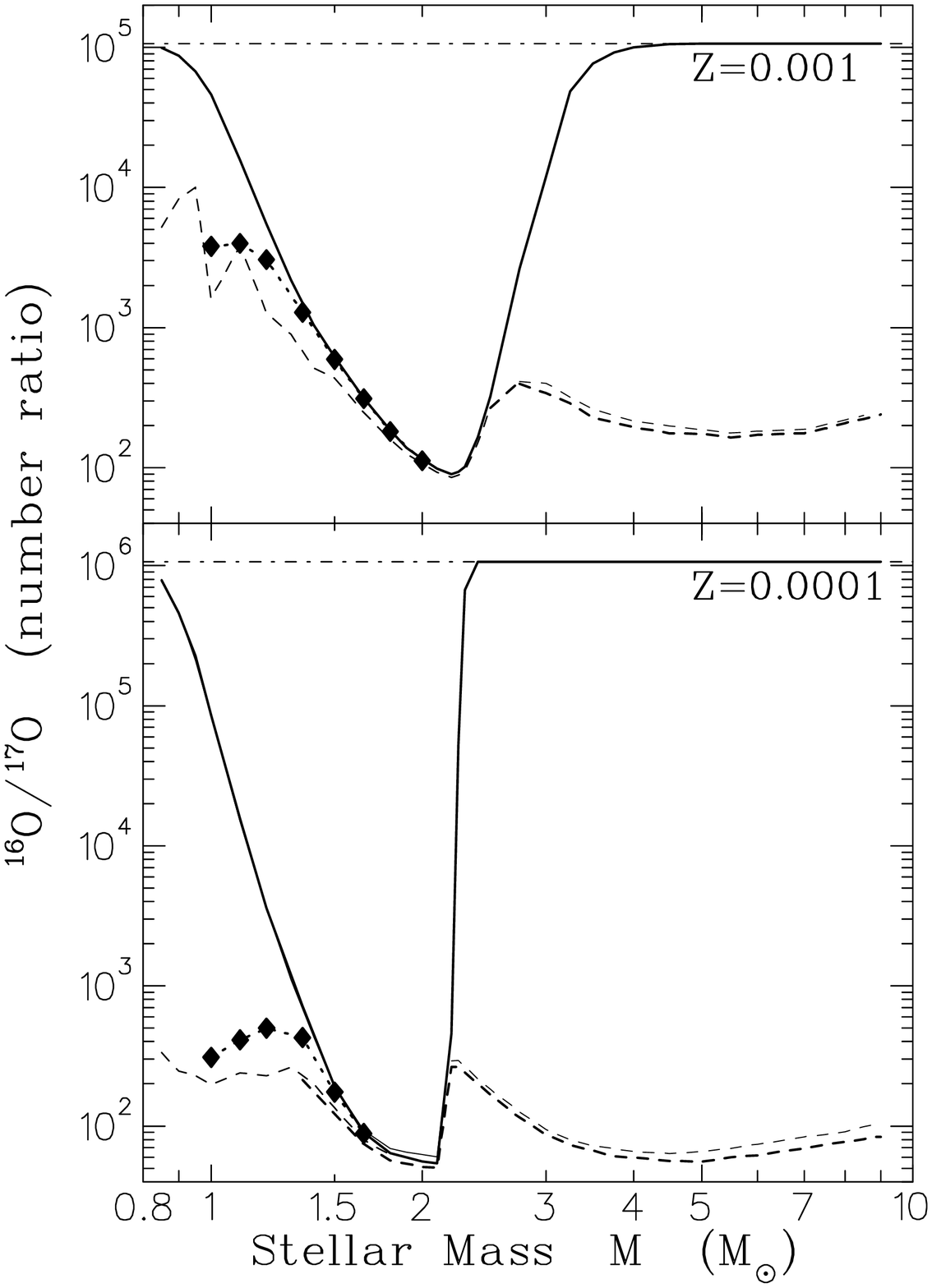}
\caption{The ${}^{16}$O/${}^{17}$O ratios resulting from first and second
dredge-up and RGB cool bottom processing; notation as in
Fig.~\protect\ref{F:lattboot:dr12cn}.  Initial stellar
${}^{16}$O/${}^{17}$O ratios were assumed to be inversely proportional
to Fe/H, as per galactic chemical evolution models of Timmes
{\it et al.}~\protect\cite{lattboot:TimWW95}.  The $\hbox{$\rm{}^{17}O$}+p$
reaction rates of Landr\'e {\it et al.}~\protect\cite{lattboot:La90} were
used in general; the slightly higher ${}^{16}$O/${}^{17}$O ratios
resulting from the more accurate Blackmon
{\it et al.}~\protect\cite{lattboot:Bla+95,lattboot:Bl96}
rates are also shown for $Z = 0.02$, 0.007, 0.001, and~0.0001
(i.e., the upper of the two solid lines, or of the two dashed lines).}
\label{F:lattboot:dr12o17}
\end{figure}

Figure~\ref{F:lattboot:dr12o17} shows the effect of first and second
dredge-up on the ${}^{16}$O/${}^{17}$O ratio.  There is no effect
for stars of $\sim 1\>M_\odot$, since the convective envelope (during
first dredge-up) does not reach the \hbox{$\rm{}^{17}O$}-pocket in
these stars.  For $1\>M_\odot \lesssim M \lesssim 2\>M_\odot$,
the convective envelope reaches partially into the
\hbox{$\rm{}^{17}O$}-pocket; since there is a steep abundance gradient at
the outer edge of this \hbox{$\rm{}^{17}O$}-pocket, the exact amount of
\hbox{$\rm{}^{17}O$} dredge-up is sensitive to the precise depth of
convection and form of the profile.  Thus there is some disagreement between
different investigators in this mass range; the
predictions for $Z = 0.02$ in Figure~\ref{F:lattboot:dr12o17}
agree well with those of Bressan {\it et al.}~\cite{lattboot:Bre+93}
and Schaller {\it et al.}~\cite{lattboot:Schal+92}, but
Dearborn~\cite{lattboot:Dea92} finds larger ${}^{17}$O-enhancements
below $1.5\>M_\odot$ (see also discussion by
El~Eid~\cite{lattboot:ElE94}).  For stars of mass $\gtrsim 2\>M_\odot$, the
convective envelope reaches down slightly below the peak of the
\hbox{$\rm{}^{17}O$} pocket, yielding large surface \hbox{$\rm{}^{17}O$}
enrichments, as shown in Figure~\ref{F:lattboot:dr12o17}.
In Population~II stars, the
\hbox{$\rm{}^{17}O$}~pocket is wider than in Population~I stars,
leading to more \hbox{$\rm{}^{17}O$} enrichment
relative to \hbox{$\rm{}^{16}O$}, although this may not be attained until
second dredge-up in higher masses, where first dredge-up may not occur
(as discussed above).
Uncertainties in the \hbox{$\rm{}^{17}O$}-destruction rates have
no effect below~$2\>M_\odot$,
because dredge-up does not reach regions where any \hbox{$\rm{}^{17}O$}
was destroyed; in this mass range, the uncertainties in the stellar
observations are too large to say more than that theory and observation are
not inconsistent~\cite{lattboot:Dea92,lattboot:ElE94,lattboot:BSW94}.
For $M \gtrsim 2\>M_\odot$,
differences between the \hbox{$\rm{}^{16}O/{}^{17}O$} results
of different investigators are largely due to use of different rates for
the \hbox{$\rm{}^{17}O$}-destruction reactions, as discussed by
El~Eid~\cite{lattboot:ElE94} (see also~\cite{lattboot:BSW94}).
The most recent $\rm{}^{17}O+p$ rates of
Blackmon {\it et al.}~\cite{lattboot:Bla+95,lattboot:Bl96}
have much smaller uncertainties; they differ by about 2-$\sigma$ (in the
``uncertain factor~$f_1$'') from the previous recommended rates of
Landr\'e {\it et al.}~\cite{lattboot:La90}, and yield abundances that differ
by less than~20\%, as shown in Figure~\ref{F:lattboot:dr12o17}
(both being consistent with observations of intermediate
mass stars~\cite{lattboot:BSW94}).  Except for stars of very low mass
($\lesssim 1\>M_\odot$), the final \hbox{$\rm{}^{16}O/{}^{17}O$} ratio is
almost independent of its initial value, as one would expect, since the
amount of \hbox{$\rm{}^{17}O$} dredged up is much
larger than the amount originally present in the envelope.

Figure~\ref{F:lattboot:dr12o17} shows that cool bottom processing on the RGB
should have little effect for Population~I stars, but should yield quite
large ${}^{17}$O~enhancements in low mass Population~II stars --- comparable
to the ${}^{17}$O~enhancements that would result from second dredge-up in
the {\it absence\/} of cool bottom processing.  Some Population~II stars
experience significant ${}^{16}$O~depletion on the RGB (this is observed in
some globular cluster stars: see above); for such stars, where the models
of the present work underestimate the extent of cool bottom processing, the
\hbox{$\rm{}^{16}O/{}^{17}O$}
ratio should approach CNO-cycle equilibrium, namely,
$100 \lesssim \rm{}^{16}O/{}^{17}O \lesssim 500$ for the relevant
H-burning temperatures.

%
\begin{figure}[t]
\lattbootplotfiddle{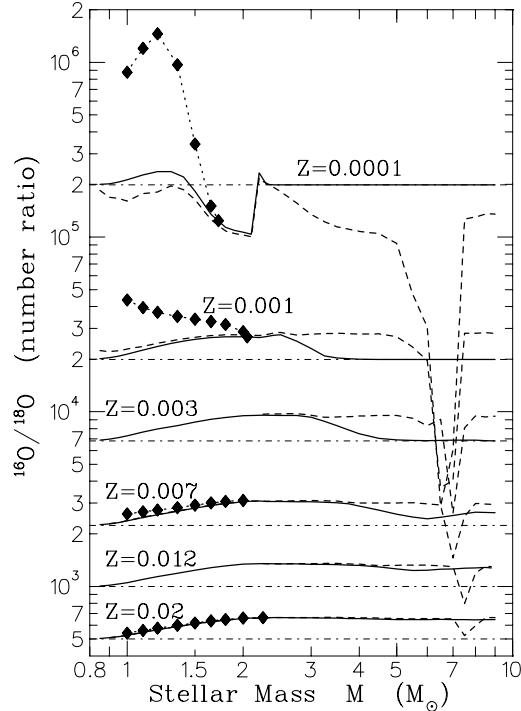}{3.5 true in}{0}{40}{40}{-130}{-26}
\caption[]{The ${}^{16}$O/${}^{18}$O ratios from first and second dredge-up
and RGB cool bottom processing; notation as in
Fig.~\protect\ref{F:lattboot:dr12cn}.  Initial stellar
${}^{16}$O/${}^{18}$O ratios were assumed to be inversely proportional
to Fe/H, as per Timmes
{\it et al.}~\protect\cite{lattboot:TimWW95}.}
\label{F:lattboot:dr12o18}
\end{figure}

Since the amount of \hbox{$\rm{}^{18}O$}-depleted matter engulfed by first
and/or second dredge-up is never a very large
fraction of the total envelope mass, dredge-up usually does not
change the surface \hbox{$\rm{}^{18}O$} abundance much, as may be seen from
Figure~\ref{F:lattboot:dr12o18}; this agrees with RGB
observations~\cite{lattboot:BSW94}.  The only exception to this is in stars
of $\sim 7\>M_\odot$, where second dredge-up may reach material containing
\hbox{$\rm{}^{18}O$} produced during core He-burning.
The {\it fractional\/} change in \hbox{$\rm{}^{16}O/{}^{18}O$} owing to first
and second dredge-up is generally almost
independent of the initial \hbox{$\rm{}^{16}O/{}^{18}O$} ratio.
Figure~\ref{F:lattboot:dr12o18} also
demonstrates that \hbox{$\rm{}^{18}O$} is essentially
unaffected by cool bottom processing in solar metallicity
stars on the RGB, but can be significantly depleted in Population~II RGB
stars of low mass.

%
\begin{figure}[t]
\lattbootplotfiddle{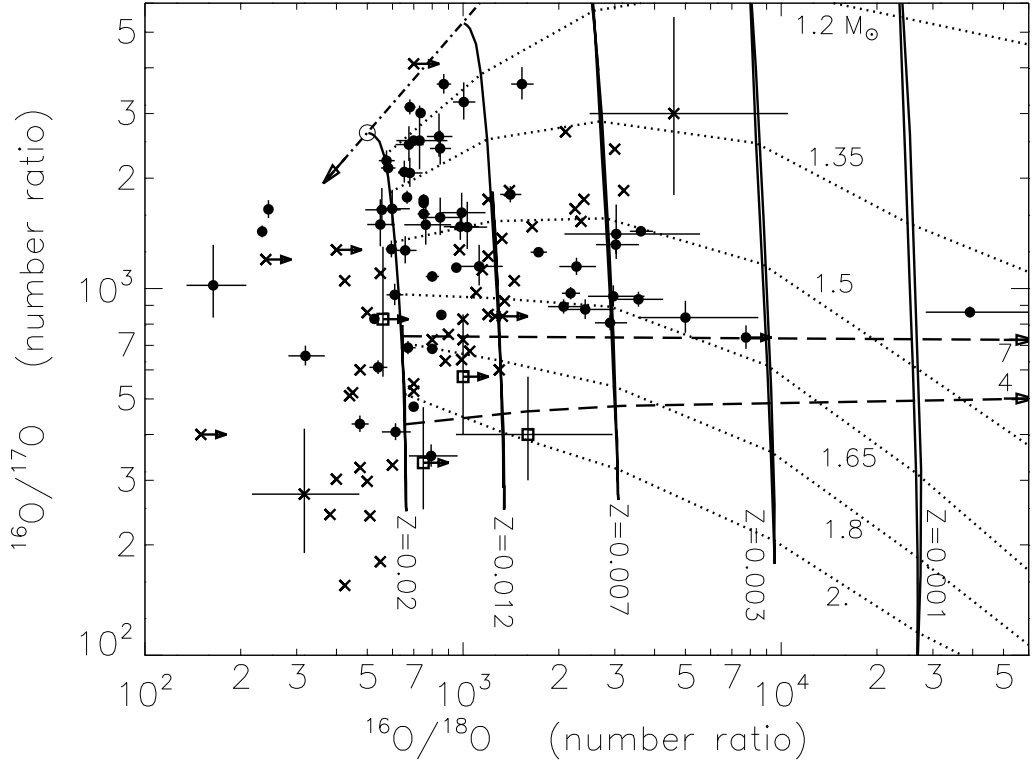}{3.8 true in}{90}{64}{64}{250}{-47}
\caption[]{The oxygen
isotope--isotope diagram.  {\it Dot-dashed line\/}: evolution of the
interstellar medium~\protect\cite{lattboot:TimWW95}
({\it open circle\/}: solar ratios).  {\it Solid lines\/} (labelled by~$Z$)
give the theoretical locus occupied by first and second dredge-up abundances.
{\it Dotted lines\/} give
first dredge-up abundances for six stellar masses (as labelled).
{\it Dashed lines\/} show the effect of TP-AGB hot bottom burning
in $Z = 0.02$ stars of 4 and $7\>M_\odot$.
{\it Crosses\/}: observed ratios in S~and C~stars on the
AGB~\protect\cite{lattboot:HarLS85,lattboot:Har+87,lattboot:Kah92}
(typical errors shown for two stars, at lower left and upper right).
{\it Open squares\/}: observed ratios in four J-type carbon stars, with
$\rm{}^{12}C/{}^{13}C \sim 3$ suggesting hot bottom burning.  {\it Solid
circles\/}: high precision grain measurements~\protect\cite{lattboot:Hus+92,%
lattboot:Hus+94,lattboot:Nit+94,lattboot:Nit97}.}
\label{F:lattboot:dr12oisoiso}
\end{figure}

While low-mass solar-metallicity stars experience little \hbox{$\rm{}^{18}O$}
depletion on the RGB, there is some observational evidence that suggests
that they may experience significant ${}^{18}$O depletion (or
${}^{16}$O enhancement) on the AGB~\cite{lattboot:WBS95,lattboot:BS97}.
Figure~\ref{F:lattboot:dr12oisoiso} shows oxygen isotope ratios observed
in AGB stars known to be in the TP-AGB phase (from the fact that they are
S~stars or C~stars, and thus must have experienced third dredge-up).  There
appears to be a trend, such that those stars whose ${}^{17}$O
abundances indicate low masses have higher \hbox{$\rm{}^{16}O/{}^{18}O$}
ratios than expected.
The indicated
\hbox{$\rm{}^{18}O$} depletions are not much larger than the observational
errors, but are exhibited by a number of stars.  One might suggest that
the trend was in fact an age effect, i.e., that low mass stars tend to be
older, and thus tend to have lower initial metallicity and lower initial
\hbox{$\rm{}^{18}O$} abundances; however, no such trend with stellar mass is
visible in the (admittedly few) RGB oxygen isotope observations.  The high
precision grain data (solid circles in Fig.~\ref{F:lattboot:dr12oisoiso})
certainly show a wide range in \hbox{$\rm{}^{18}O$} abundances.  Only four
of the most \hbox{$\rm{}^{18}O$}-depleted grains
have abundances consistent with an
origin in a $7\>M_\odot$ star that is undergoing hot bottom burning on
the AGB (see section~\ref{S:lattboot:HBB} and dashed lines in
Fig.~\ref{F:lattboot:dr12oisoiso}).  To explain the rest by variations
in the initial isotope
ratios would require that these grains originated in $\sim 1.6\>M_\odot$ stars
with metallicity $\sim 1/3$ of solar --- not impossible, but not what
one would have expected, since most of the other
grains exhibit \hbox{$\rm{}^{16}O/{}^{18}O$} ratios indicative of
metallicities $> 2/3$~of solar.  It has been suggested that
\hbox{$\rm{}^{18}O$} depletions in low mass AGB stars would occur naturally
if the extra mixing/cool bottom processing mechanism operated in at least
some AGB stars~\cite{lattboot:BSW95,lattboot:WBS95} (as well as RGB stars).
If this were the explanation for the AGB \hbox{$\rm{}^{18}O$} depletions,
then the extra mixing on the AGB would have to reach significantly deeper
into the H-burning shell than was the case on the RGB, in order to
yield significant \hbox{$\rm{}^{18}O$} depletions on the short AGB
timescale.  Such stars would be expected to have
${}^{12}$C/${}^{13}$C ratios not far above the nuclear equilibrium value
of~3, and to convert significant amounts of ${}^{12}$C into~${}^{14}$N,
yielding nitrogen enrichments and making it harder to become a carbon
star;
stellar observations show little or no
indication of these other consequences of AGB cool bottom processing,
though there is insufficient data to rule out AGB cool bottom processing.
An alternative explanation is suggested by a parameterized
convective overshoot model of Herwig {\it et al.}~\cite{lattboot:HarLS85},
which suggested that third dredge-up could yield significant enrichment
of envelope~${}^{16}$O (in contrast to standard mixing models); if this
were the case, the ${}^{16}$O$/{}^{18}$O ratio might increase (by a factor
of~$\lesssim 2$) due to third
dredge-up (the ${}^{16}$O$/{}^{17}$O ratio would increase slightly less,
as some ${}^{17}$O is produced in the H-shell).

\section{Thermally Pulsing-AGB Evolution}  \label{S:lattboot:TPAGB}

%
\begin{figure}[t]
\lattbootplotfiddle{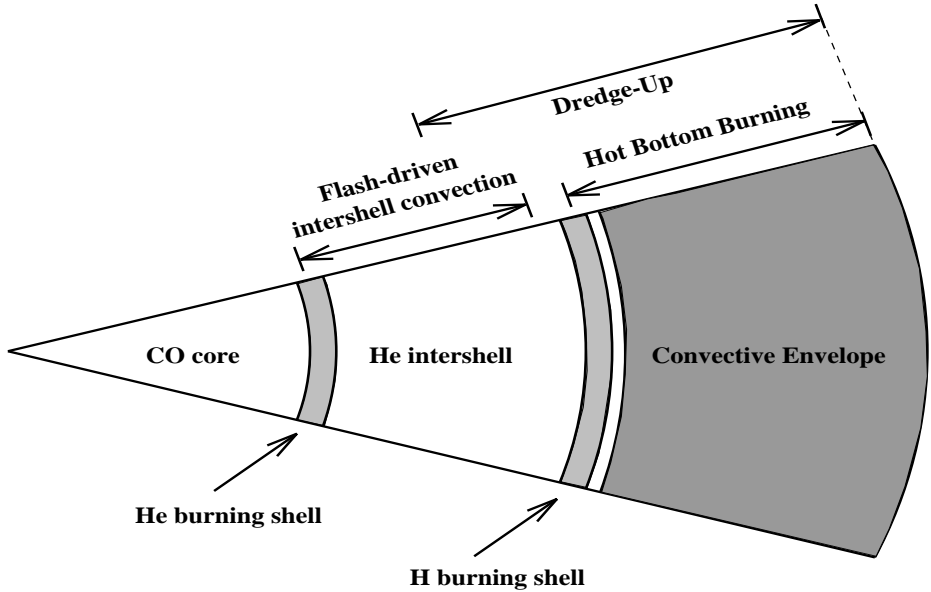}{2.6 true in}{0}{55}{48}{-170}{-20}
\vskip 0.4 true in
\caption{Schematic structure of an AGB star ({\bf not} to scale). 
During a thermal pulse
the intershell convection extends over the region marked by
``flash-driven intershell convection''. During the
dredge-up phase, the convective envelope moves inward to the depth marked
by ``dredge-up''. For massive stars, during the interpulse phase
the convective envelope penetrates the top of the H-shell,
so that envelope convection extends down to the region marked ``hot bottom
burning''.}
\label{F:lattboot:AGBstruc}
\end{figure}

The details of AGB evolution have been the subject of much theoretical work,
and are, in one sense, quite well understood. The reader is referred 
to~\cite{lattboot:IR83,lattboot:FL96a,lattboot:Letal96} 
for details. In 1981, Renzini \& Voli~\cite{lattboot:RenV81} attempted to
combine theory and observations into a consistent set of nucleosynthetic
yields for low and intermediate mass AGB stars, using parameterized
``synthetic'' AGB nucleosynthesis models.  Recently, several authors have
used more extensive observations and updated stellar evolution models in
a similar manner, to provide improved sets of AGB nucleosynthetic
yields~\cite{lattboot:MarBC96,lattboot:vdHG97,lattboot:ForC97}.  As discussed
below, there are still sufficient uncertainties in AGB evolution that such
calculated yields should be taken with a grain of salt, particularly for
the elements heavier than oxygen, where nuclear rate uncertainties can
be considerable~\cite{lattboot:ArnMC95}.

Briefly, an AGB star has the
structure shown in Figure~\ref{F:lattboot:AGBstruc}.
The C-O core is the result of He burning, and will become the final
white-dwarf remnant. Just above this is the He-shell. This is thermally
unstable, and burns vigorously during shell flashes (or thermal pulses)
but is essentially extinguished between them. Above the He-shell is the
intershell region, so-called because it is between the He and H-shells.
Above the H-shell is the convective envelope.
During a thermal pulse, the He-shell will deliver some $10^7\>L_\odot$
for a brief period, and this enormous energy production results in the
formation of a
convective zone. This ``flash-driven intershell convection'' extends
over the region shown in Figure~\ref{F:lattboot:AGBstruc}, and thus 
mixes the products of (partial)
He-burning throughout this region. The approximate composition of this
zone is 25\%~${}^{12}$C and 75\%~${}^4$He (note, however, that a parameterized
convective overshoot model of Herwig {\it et al.}~\cite{lattboot:HarLS85}
has yielded a composition 50\%~${}^{12}$C, 25\%~${}^{16}$O,
and 25\%~${}^4$He, suggesting that uncertainties in convective mixing can have
a significant effect on flash nucleosynthesis: see
section~\ref{S:lattboot:TDU}).  There are also significant
overabundances of~${}^{22}$Ne, produced via
${}^{14}$N$(\alpha,\gamma)^{18}$O$(\alpha,\gamma)^{22}$Ne (note that
the H-burning shell has converted almost all the CNO elements
into~${}^{14}$N), and of the results of neutron-capture nucleosynthesis,
namely, ${}^{19}$F and ``$s$-process isotopes''; ${}^{19}$F~is produced when
protons from $(n,p)$ reactions allow the reaction pathway
${}^{18}$O$(p,\alpha)^{15}$N$(\alpha,\gamma)^{19}$F to
proceed~\cite{lattboot:For+92,lattboot:Metal96,lattboot:ForC97,%
lattboot:Was+94}.  Following the pulse, the
helium luminosity decreases and the star expands. This essentially
extinguishes the H-shell, and the bottom of the convective
envelope moves inwards in mass. After a small number of pulses, this
convection penetrates the region that was mixed by
the flash-driven convective zone; this results
in the mixing of freshly produced carbon to the stellar surface
(``{\it third dredge-up\/}''). As the star 
begins to contract back to its normal configuration, the H-shell
is re-ignited and provides all of the energy during the next interpulse
phase, until the following pulse.

For more massive stars (above $\sim 4\>M_\odot$) the bottom of
the convective envelope penetrates into the top of the H-shell,
and some nuclear reactions take place at the bottom of the
convective envelope. This is known as ``hot bottom burning'' (hereafter HBB),
and is shown schematically in Figure~\ref{F:lattboot:AGBstruc}.
The AGB phase terminates when
mass loss has reduced the star's envelope mass almost to zero.
In the calculations presented below we used the mass loss
formula of Vassiliadis \& Wood~\cite{lattboot:VW93}, although there are
other formulations and the number of thermal pulses, HBB nucleosynthesis,
and final stellar mass depend sensitively on this rather uncertain input (see
also~\cite{lattboot:BoSaAh93,lattboot:Bloc95,lattboot:ForC97,%
lattboot:Stran+97}).  The other
main sources of uncertainty in AGB evolution are possible errors in
low-temperature molecular
opacities~\cite{lattboot:SB91,lattboot:BS92,lattboot:BoSaAh93},
and uncertainties in the
treatment of convective mixing (see section~\ref{S:lattboot:TDU}).

%
\begin{figure}
\lattbootplotfiddle{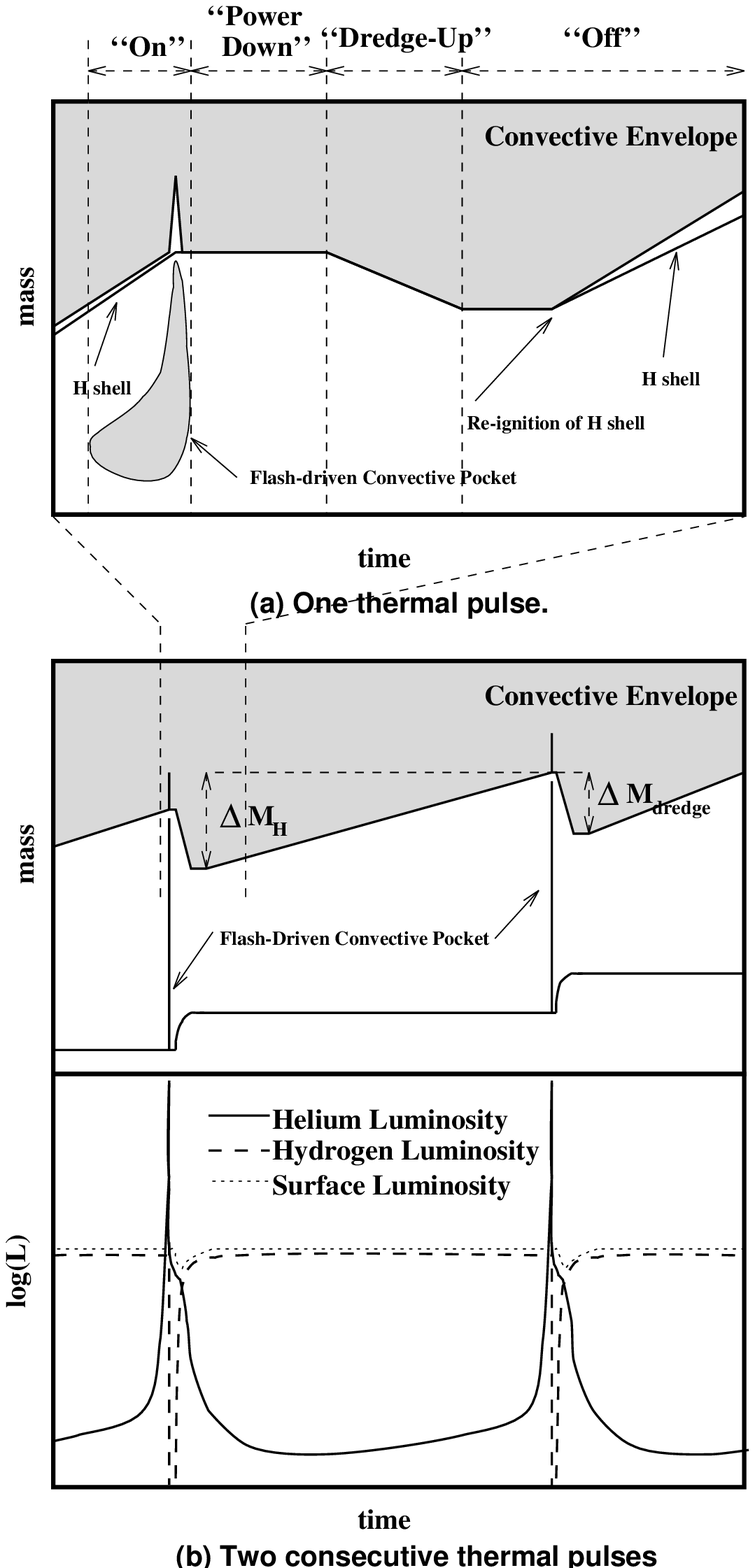}{8.0 true in}{0}{95}{82}{-150}{65}
\vskip -0.71 true in
\caption{The four phases of a thermal pulse. In figure~(a) we see the
detailed evolution during a thermal pulse, showing the intershell 
convection zone and the dredge-up which follows the pulse. Figure~(b)
shows two consecutive pulses, and defines the two masses $\Delta M_H$ and
$\Delta M_{dredge}$. These define the dredge-up parameter $\lambda = 
\Delta M_{dredge}/\Delta M_H$. The bottom panel shows the typical luminosity
variations during a flash cycle.}
\label{F:lattboot:TwoTP}
\end{figure}

The repeated operation of the thermal pulse cycle described above is
responsible for the periodic addition of carbon to the
stellar surface. A crucial parameter in this evolution is the
amount of dredge-up, as measured by the so-called ``dredge-up parameter''
$\lambda = \Delta M_{dredge} / \Delta M_{H}$,
where
$\Delta M_{dredge}$ is the amount of matter dredged-up following a given pulse,
and $\Delta M_H$ is the amount of matter processed by the H-shell 
between pulses. These parameters are defined in Figure~\ref{F:lattboot:TwoTP}.
If enough pulses occur (with sufficient dredge-up
per pulse) eventually the star becomes a Carbon star, with $n($C$) > n($O$)$.
These stars are very important for the interpretation of meteoritic grains
because it is believed that most SiC grains form in their carbon-rich
envelopes.
But this production of carbon is really on the tip of the nucleosynthetic
ice-berg! We shall deal with some of the other nucleosynthesis products
below. For information about the production of fluorine, however, please
refer to~\cite{lattboot:Metal96,lattboot:ForC97}.

%
\begin{figure}
\lattbootplotfiddle{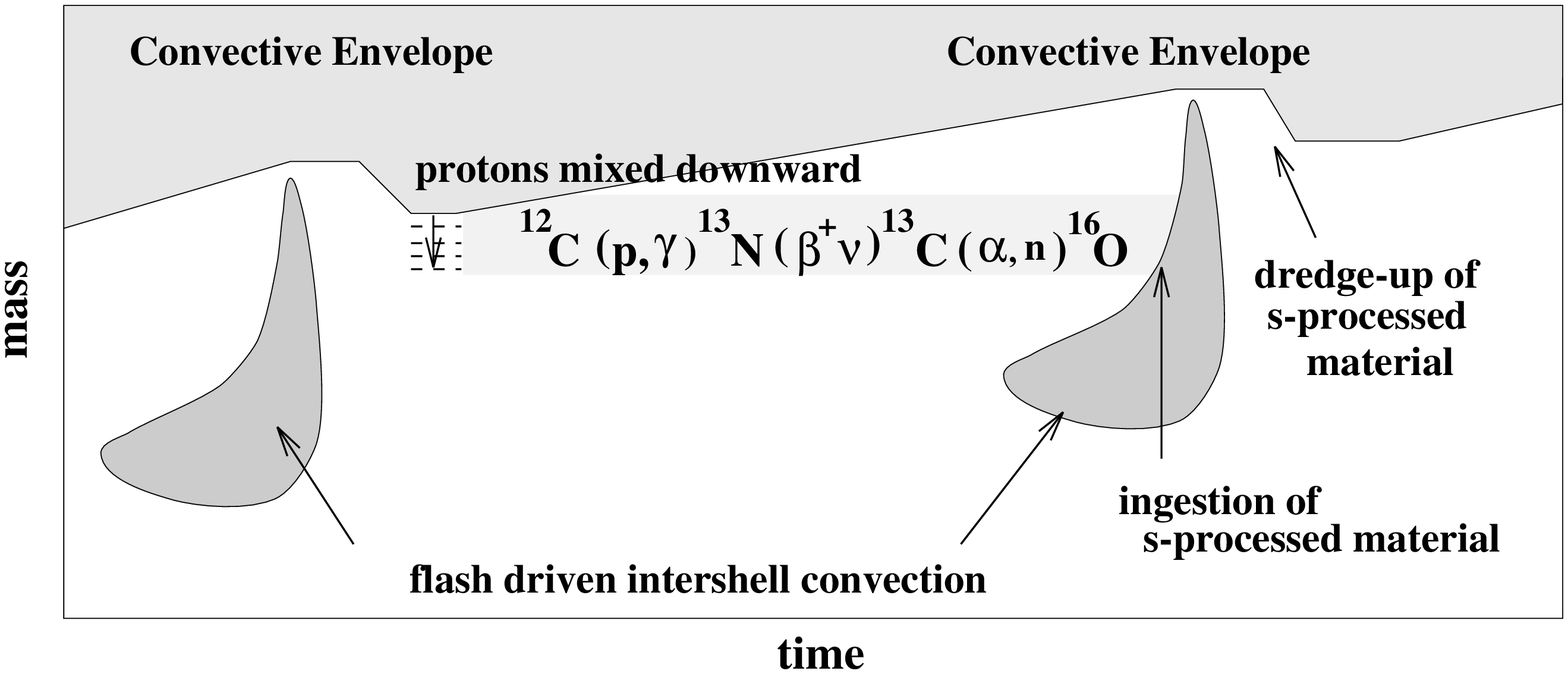}{2.5 true in}{0}{50}{60}{-190}{0}
\caption{Schematic structure of two consecutive thermal pulses,
showing how the downward mixing of hydrogen leads to the production
of $^{13}$C and then $s$-processing. Note that this
$s$-processed material is later dredged to the surface of the star.}
\label{F:lattboot:C13pocket}
\end{figure}

Neutron capture nucleosynthesis
in AGB stars is well known, both
observationally and theoretically. We now believe that the neutron source
active in these stars (at least the lower masses) is $^{13}$C. But how
exactly does this $^{13}$C arise?
This remains a serious problem for the models, which do not exhibit
the required $^{13}$C production.
While it is true that some $^{13}$C is produced by CNO cycling
in the H-shell, this is nowhere near enough to produce the neutron
exposures inferred from observations of these stars.
We will make an {\it ad hoc\/} assumption that some kind of
extra mixing takes place at the
bottom of the convective envelope during the dredge-up phase. This has
indeed been found in one of the models of Iben \&
Renzini~\cite{lattboot:IR82a,lattboot:IR82b}, but has not been reproduced
by their subsequent models~\cite{lattboot:Iben83},
nor by the models of other authors.  However, such partial mixing at
the base of the envelope convective arises naturally from hydrodynamic
simulations of overshoot below a convective region~\cite{lattboot:Hurl+94,%
lattboot:RieuZ95,lattboot:FreyLS96} (and with a parameterization
of such an overshoot model, Herwig {\it et al.}~\cite{lattboot:Herw+97} did
indeed find that a ${}^{13}$C-pocket was produced).
Assuming that such mixing does indeed occur, small amounts of hydrogen are
mixed into a region which is relatively rich in $^{12}$C. During the
subsequent interpulse phase, these regions heat and the protons are captured
by the $^{12}$C to produce $^{13}$C. (It is crucial that there is not too
much hydrogen in this region, or all the $^{13}$C will be burned
to $^{14}$N\hbox{}.)
This ``$^{13}$C-pocket'' was initially believed to sit in the star and 
wait until the next thermal pulse, when it would be
engulfed by the intershell convection. The high temperatures present
would then release the neutrons and the $s$-processing would occur in the
convective intershell region. But it has been shown recently
(Straniero {\it et al.}~\cite{lattboot:Setal95}, followed
by~\cite{lattboot:Letal96,lattboot:Metal96}) that
the temperatures in the intershell are sufficiently high (i.e., the interpulse
duration is sufficiently long) that almost all of
the $^{13}$C burns to $^{16}$O there, releasing the neutrons locally and
thus resulting in $s$-processing {\it in situ\/} with large neutron
exposures.  Later, when this region is mixed into the flash-driven
convection zone, the results of the neutron captures are also mixed into
the convective zone. This situation is shown schematically in 
Figure~\ref{F:lattboot:C13pocket}. A brief pulse of neutrons from the
${}^{22}$Ne neutron source, at the peak of the shell flash,
may redistribute $s$-process yields somewhat in the flash-driven convection
zone.  The details of the $s$-processing are covered in the excellent
review by Gallino \& Busso~\cite{lattboot:GBhere}; $s$-process models
(parameterized by the size of the ${}^{13}$C-pocket) can
yield a good match with observed $s$-process abundances.

\section{The Third Dredge-Up} \label{S:lattboot:TDU}

The nucleosynthesis described above results in changes in the photospheric
composition because of the operation of the third dredge-up. Yet there are
several details concerning this mechanism which are not well understood.
One usually applies the Schwarzschild criterion to determine
the convective boundary. This relies on finding the position where 
the acceleration of gas eddies
is zero, which is where $\nabla_{rad} = \nabla_{ad}$.
Convective eddies will still have a non-zero
momentum when they reach this boundary, and hence they will penetrate
into the radiatively stable region where they are decelerated to zero velocity
(``{\it convective overshoot\/}''). But for third dredge-up,
the boundary is even more prone to mixing.
The minimum value of $\nabla_{rad}/\nabla_{ad}$
(at the bottom of the convective envelope) exceeds unity substantially, and
hence a finite buoyancy will drive eddies into the stable region.
Exactly how the star will mix, to achieve
the expected convective neutrality, is uncertain. It is, of course, a
hydrodynamical problem. We expect the convective region to grow
into the intershell region, until the gradients smoothly approach each
other. From this configuration we would still expect the usual overshoot.

This situation has been investigated
in some detail by~\cite{lattboot:FL96b} who found that the depth of
dredge-up depends critically on assumptions made at the boundary
of the convective region, as well as on the way in which the mixing
is handled within the evolutionary calculation (e.g., if the mixing
is performed after each iteration, or only after a model has converged);
increasing the number of mass zones and time steps in the model can also have
a significant effect~\cite{lattboot:Stran+97}.
Current work in progress has shown that the depth of dredge-up dramatically
alters the evolution of the star. Deep dredge-up cools the intershell region,
and slows the advance of the He-shell to almost zero, yielding almost
stationary shell burning.
The depth of
the dredge-up depends also on the treatment of the entropy cost of mixing
dense material upward in a gravitational field~\cite{lattboot:W81}.
Hydrodynamic simulations of stellar convection in 2-D and 3-D have been used
in attempts to parameterize the extent of overshoot below a convective
envelope that results from the downward plumes typically found in such
simulations (e.g.,~\cite{lattboot:Hurl+94,lattboot:RieuZ95,lattboot:FreyLS96}).
Herwig {\it et al.}~\cite{lattboot:Herw+97} applied the
overshoot parameterization of~\cite{lattboot:FreyLS96} to all boundaries of
all convective regions, finding significant effects on thermal pulse
nucleosynthesis and dredge-up.  We have much to learn about this
complicated phase of evolution, and work is continuing.    

\section{Hot Bottom Burning}  \label{S:lattboot:HBB}

Hot Bottom Burning (hereafter ``HBB'') is the colourful name
given to the circumstance where
the temperature at the bottom of a star's convective envelope is sufficiently
high for  nucleosynthesis to take place. It is perhaps better
to think of this as the bottom of the convective envelope penetrating into the
top of the H-burning shell.
It is also
sometimes known as ``convective envelope burning'' or simply ``envelope
burning''. 
The first calculations of HBB were performed by
Sackmann {\it et al.}~\cite{lattboot:SSD74} and
Scalo {\it et al.}~\cite{lattboot:SDU75};
for a discussion of the development and history of this
phenomenon, please see~\cite{lattboot:Letal97}. 

It is only relatively recently that stellar models have shown deep
convective envelopes with temperatures exceeding 80 million 
degrees\cite{lattboot:BlSch91,lattboot:L92} in a thin region 
at their base. Indeed, it was noticed by\cite{lattboot:BlSch91} that this
resulted in the star no longer obeying the \hbox{core-mass} {\it vs.}\
luminosity relation, but space prevents us from going into details here (see 
\cite{lattboot:BlSch91,lattboot:BS92,lattboot:Letal97} for more information).

The first effect of HBB is the production of $^7$Li via the Cameron-Fowler
Beryllium Transport Mechanism.
The first systematic studies of this were
carried out by~\cite{lattboot:SackBoot92}. The key ingredient is an algorithm
for time-dependent mixing because it is essential that the timescale for
mixing be much shorter than the electron-capture lifetime
for $^7$Be (instantaneous mixing would decrease~$^7$Li!).
Sackmann \& Boothroyd~\cite{lattboot:SackBoot92} showed that $^7$Li
was produced when the temperature~$T_{bce}$ at the bottom of the
convective envelope exceeded $50\times 10^6$K, with abundances
up to $\log\varepsilon(^7$Li$) \sim 4.5$ in stars with
$M_{bol} \simeq -6$ to~$-7$ (note that
$\log\varepsilon(^7$Li$) \equiv \log\{n(^7$Li$)/n($H$)\}+12$); similar
results (with slightly higher peak lithium abundances) were obtained in HBB
models of Forestini \& Charbonnel~\cite{lattboot:ForC97}.
This is in excellent agreement with the
observations~\cite{lattboot:SL89,lattboot:SL90,lattboot:Abia+91,%
lattboot:Smithetal95} which show Magellanic Cloud AGB
stars of $M_{bol}$ between $-6$ and $-7$
to have $\log\varepsilon(^7$Li$)$ in the range $2.2$ to~$3.8$, and
galactic AGB stars with $\log\varepsilon(^7$Li$) \lesssim 5$. 

One of the main effects of HBB is to take the $^{12}$C which is
dredged to the surface and process it into $^{14}$N via the CN cycle.
Some $^{13}$C will also be produced, but the overall
carbon destruction prevents the star from becoming a carbon star.
Detailed calculations of the effect of HBB on CNO elements were
carried out by Boothroyd {\it et al.}~\cite{lattboot:BoSaAh93}
(see also~\cite{lattboot:MarBC96,lattboot:ForC97,lattboot:Letal97}).
Significant destruction of $^{12}$C together with production of $^{13}$C and
$^{14}$N requires temperatures of at least $80\times 10^6$K\hbox{}.
This was found for masses greater than $\sim 3.5$ to~$4\>M_\odot$,
depending on the metallicity
(increasing as $Z$ increases). These authors expect
a maximum luminosity of $M_{bol} \approx -6.4$ for carbon stars,
as higher luminosities
will result in HBB processing the $^{12}$C into~$^{14}$N.
Despite the narrowness of the burning layer, the mixing
timescale is such that the entire envelope is processed
through the burning region
many times during the interpulse phase, and the nuclear equilibrium
ratio $^{12}$C$/^{13}$C${} \approx 3$ (by number) is reached
for stars with $M_{bol} \gtrsim -6.3$. Thus there is a narrow region
(in luminosity) where stars
may be rich in~$^{13}$C, prior to further
processing of~$^{12}$C (and~$^{13}$C)
into~$^{14}$N.   

The situation with oxygen isotopes is a little more complicated. The
third dredge-up appears to have a negligible effect on the oxygen isotopes,
but HBB
initially destroys any $^{18}$O which is present, and then follows this by
producing some $^{17}$O~\cite{lattboot:BSW95}.
The complication comes from the fact that many AGB stars do not seem to
fit this
pattern. This is despite the fact that extremely precise meteoritic grain
analysis~\cite{lattboot:Nit+94}
reveals isotopic ratios which do fit the
results of standard first and second dredge-up models quite well (see
\cite{lattboot:BSW94}).
It now appears that to match the observations of S and C stars we may need
to invoke some
form of ``deep extra mixing'' during the AGB evolution (possibly in
addition to the first ascent of the giant branch, as discussed in
section~\ref{S:lattboot:dr12}). 

We mentioned above $^{19}$F enrichment from third dredge-up.
If a star is massive enough for HBB to develop, then any
$^{19}$F added to its envelope is destroyed by 
$^{19}$F$(p,\alpha)^{16}$O\hbox{}. Thus stars with enhanced $^{19}$F are
presumably lower-mass stars, where HBB does not take place.  This is
consistent with the fact that they are often carbon stars too.
But more can be learned from~$^{19}$F\hbox{}.        
To produce the largest observed enhancements of~$^{19}$F
seems to require a substantial source
of~$^{13}$C~\cite{lattboot:Metal96,lattboot:ForC97}, 
just as is needed for the $s$-process abundances.

Of particular interest to us at this meeting is the production
of $^{26\!}$Al in AGB stars. During hot H-burning there is some
production of $^{26\!}$Al from the Mg-Al reaction chains. This was studied 
by~\cite{lattboot:Fetal91,lattboot:Was+94,lattboot:ForC97} who looked into
the dredge-up of any $^{26\!}$Al produced by the H-shell.
They found that enhancements of $^{26\!}$Al could occur, with
$^{26\!}$Al/$^{27\!}$Al${} \sim 0.001 - 0.01$, reproducing
the mean observed isotopic ratios.
But to obtain the large enhancements required by some grains
would require dredge-up to occur when the stellar envelope mass was very small
(thus minimizing the dilution of the dredged-up material).
Yet the dredge-up stops when the envelope mass decreases too much (the critical
envelope mass for dredge-up is very uncertain). An alternative scenario is that
HBB will produce $^{26\!}$Al in the stellar envelope during the interpulse 
period.

\begin{figure}
\lattbootplotfiddle{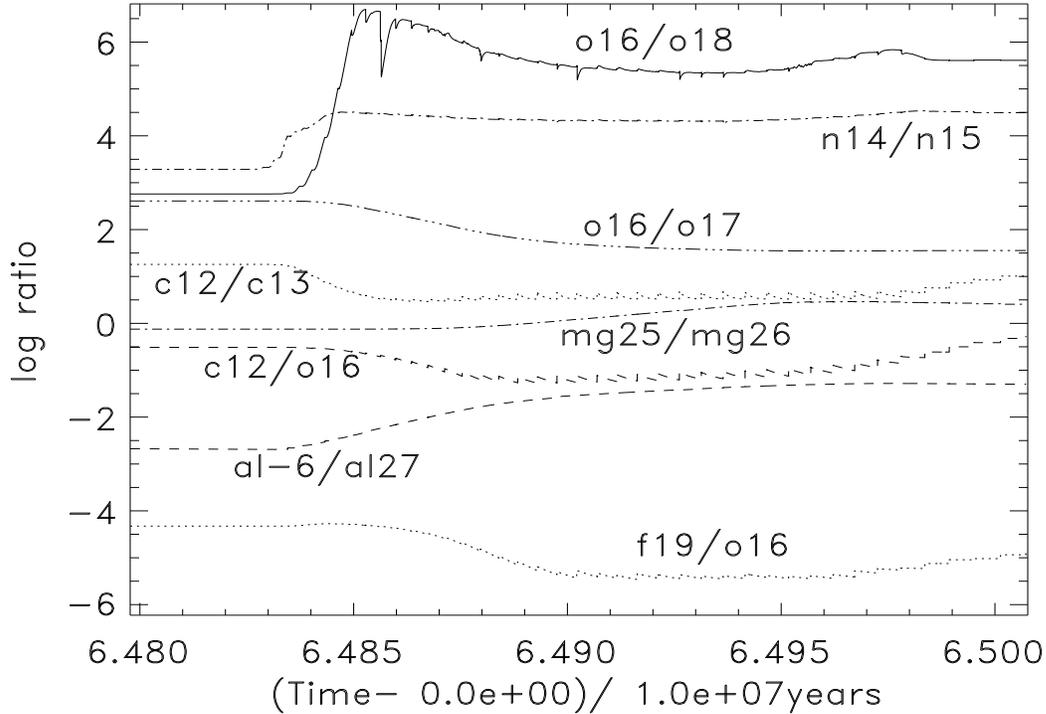}{3.0 true in}{90}{65}{60}{240}{-80}
\vskip 0.6 true in
\caption{Surface ratios during the AGB evolution of a $6\>M_\odot$ model
with $Z=0.02$.}
\label{F:lattboot:HBBSolar}
\end{figure}

We close now with some preliminary calculations of
HBB from~\cite{lattboot:Letal97}. The calculations are for $6\>M_\odot$ models
with three compositions, appropriate to the Sun ($Z=0.02$), the Large 
Magellanic Cloud ($Z=0.008$)
and the Small Magellanic Cloud ($Z=0.004$). Figure~\ref{F:lattboot:HBBSolar}
shows the results for $Z=0.02$ (in these figures, ``al-6'' represents 
the ground state of $^{26\!}$Al).
The Mg-Al chain is producing $^{26\!}$Al and $^{25}$Mg is being produced by
$^{22}$Ne$(\alpha$,n$)$ reactions in the intershell convective zone, which is
dredged to the surface after each pulse.
Hence the ratios  $^{26\!}$Al/$^{27\!}$Al and $^{25}$Mg/$^{26}$Mg
both increase. 
We already see that the high $^{26\!}$Al/$^{27\!}$Al
ratio of up to 0.05 is in good agreement with the most $^{26\!}$Al-rich
meteorite grains (see the chapters by Hoppe \& Ott and Nittler in
this volume).  Unfortunately, order-of-magnitude uncertainties in some
magnesium and aluminum burning reactions yield correspondingly large
uncertainties in the resulting abundances of ${}^{26}$Mg, ${}^{27\!}$Al,
and particularly in~${}^{26\!}$Al~\cite{lattboot:ArnMC95}.

%

%
\begin{figure}
\vskip 0.8 true in
\lattbootplotfiddle{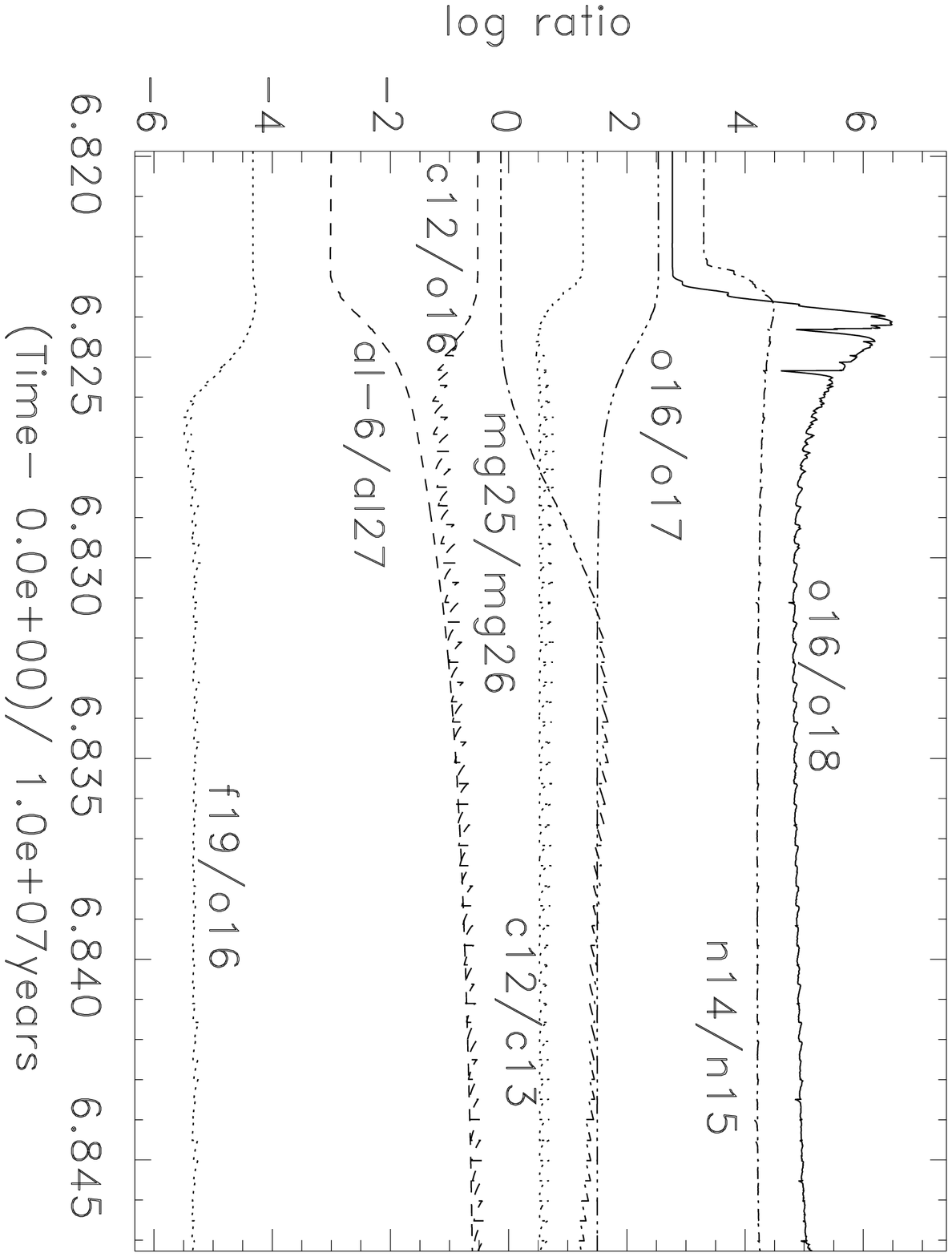}{3.0 true in}{90}{65}{60}{240}{-20}
\vskip -0.3 true in
\caption{Surface ratios during the AGB evolution of a $6\>M_\odot$ model
with $Z=0.008$.}
\label{F:lattboot:HBBLMC}
\vskip 1.5 true in
\lattbootplotfiddle{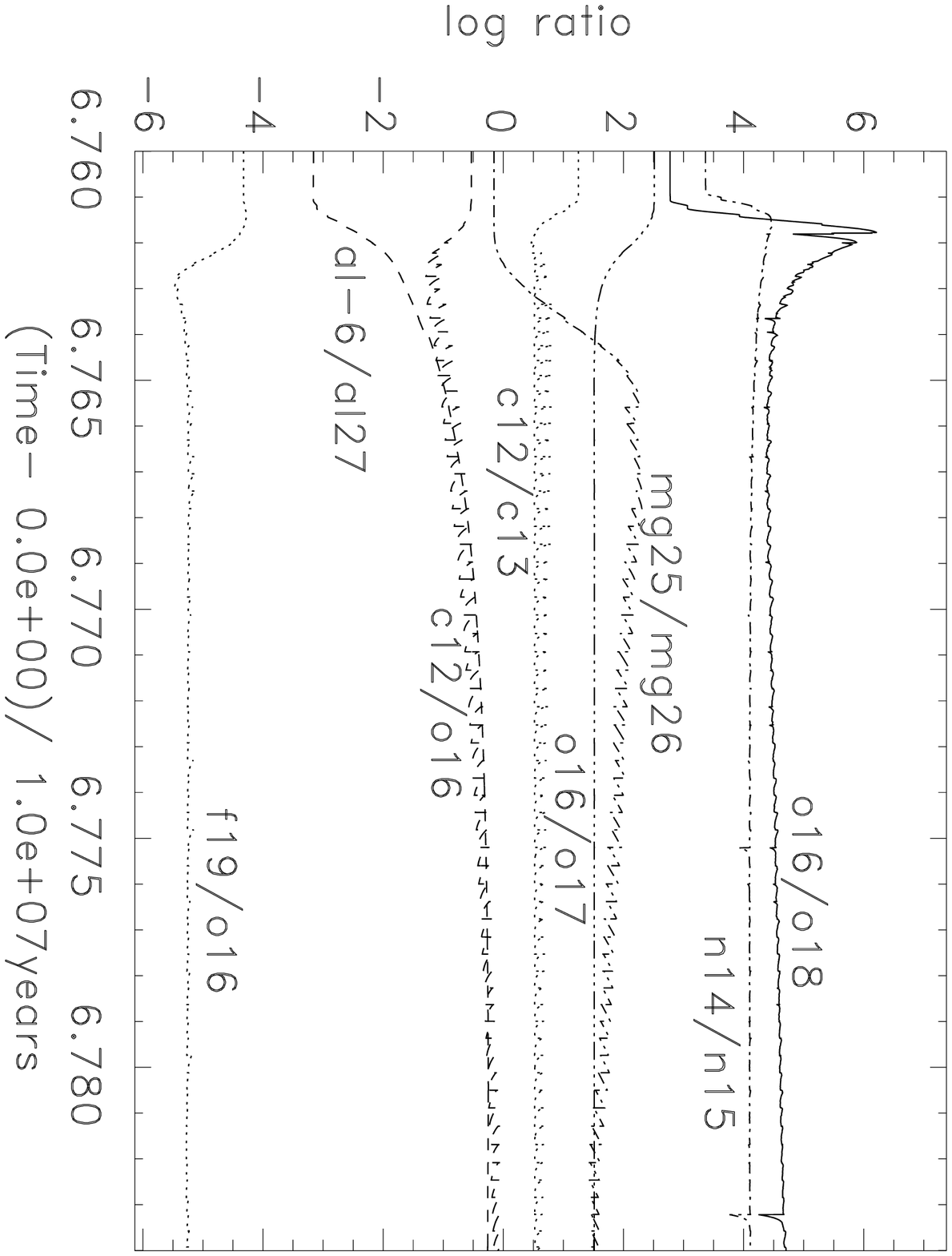}{3.0 true in}{90}{65}{60}{240}{0}
\vskip -0.5 true in
\caption{Surface ratios during the AGB evolution of a $6\>M_\odot$ model
with $Z=0.004$.}
\label{F:lattboot:HBBSMC}
\end{figure}

As the envelope mass decreases from mass loss
the HBB is shut down but third dredge-up continues. The details of the
surface composition in the latter stages of a star's life depend critically
on the competition between these effects (e.g.,~\cite{lattboot:GdJ93}).
We find that dredge-up is still strong at the 43rd
pulse (when
the mass has been reduced to $2.4\>M_\odot$) when we stopped calculations.
Figure~\ref{F:lattboot:HBBSolar} shows that the C/O ratio is kept
below unity (by HBB) but
begins to increase again from dredge-up of $^{12}$C once the HBB ceases. 
Likewise,
the $^{12}$C/$^{13}$C\ and $^{14}$N/$^{15}$N\  ratios begin to deviate from
equilibrium once HBB ends.
Although HBB has prevented the model from being a carbon star for most of its
lifetime on the AGB,
the continuing third dredge-up may yet produce a
carbon star, but now rich in $^{13}$C.
We shall address this point in
a later paper.

Figure~\ref{F:lattboot:HBBLMC} shows the surface composition for the  
$Z=0.008$ model.
Dredge-up and HBB are still operating at the end of the calculations
shown. Note that 
$^{26\!}$Al/$^{27\!}$Al and $^{25}$Mg/$^{26}$Mg
are substantially higher than found for $Z=0.02$. 
The ratio $^{12}$C/$^{13}$C
is still in equilibrium, although the deep dredge-up is continuing to
increase the C/O ratio despite HBB\hbox{}. This star may also become a
$^{13}$C-rich carbon star.                 
The surface compositions of the $Z=0.004$ model are shown in 
Figure~\ref{F:lattboot:HBBSMC}.
The trends seen in Figure~\ref{F:lattboot:HBBLMC} continue here. 
The model has essentially reached
C/O = 1, despite the fact that HBB is still operating, and producing
a large amount of $^{13}$C\hbox{}. Note also that we find
$^{26\!}$Al/$^{27\!}$Al${} \simeq 0.6$.

\section{Conclusions}  \label{S:lattboot:C}

In the last decade we have made enormous progress in our understanding of
nucleosynthesis and mixing in RGB and AGB
stars, and are beginning to understand the origin of pre-solar grains.
The rich variety of observations and measurements seem to fit the qualitative
pictures described in this paper. Quantitative results are in
agreement for some cases, but not all. Our understanding of the
processes in the stellar interior that control nucleosynthesis and
dredge-up still has several crucial gaps, which will require a great deal
of work by experimentalists, observers, and theoreticians to fill in.


\begin{references}

\bibitem{lattboot:Abia+91} Abia,~C., Boffin,~H.~M.~J., Isern,~J., \&
  Rebolo,~R., {\it Astron.\ Astrophys.\ (Lett.)}, {\bf 245}, L1 (1991).
\bibitem{lattboot:ArnMC95} Arnould,~M., Mowlavi,~N., \& Champagne,~A., in
  {\it Stellar Evolution: What Should Be Done?}, Proc.\ of 32nd 
  Li\`ege Colloquium, ed.\ A.~Noels, D.~\hbox{Frepont-Caro}, M.~Gabriel,
  N.~Grevesse, \& P.~Demarque (U.~de Li\`ege: Belgium), p.~17 (1996).
\bibitem{lattboot:Bl96} Blackmon,~J.~C., private communication (1996).
\bibitem{lattboot:Bla+95} Blackmon,~J.~C., Champagne,~A.~E., Hofstee,~M.~A.,
  Smith,~M.~S., Downing,~R.~G., \& Lamaze,~G.~P.,
  {\it Phys.\ Rev.\ Lett.}, {\bf 74}, 2642 (1995).
\bibitem{lattboot:Bloc95} Bl\"ocker,~T., {\it Astron.\ Astrophys.},
  {\bf 297}, 727 (1995).  
\bibitem{lattboot:BlSch91} Bl\"ocker,~T., \& Sch\"onberner,~D.,
  {\it Astron.\ Astrophys.\ (Lett.)}, {\bf 244}, L43 (1991).
\bibitem{lattboot:BS88} Boothroyd,~A.~I., \& Sackmann,~\hbox{I.-J.},
  {\it Astrophys.~J.}, {\bf 328}, 653 (1988).
\bibitem{lattboot:BS92} Boothroyd,~A.~I., \& Sackmann,~\hbox{I.-J.},
  {\it Astrophys.~J. (Lett.)}, {\bf 393}, L21 (1992).
\bibitem{lattboot:BS97} Boothroyd,~A.~I., \& Sackmann,~\hbox{I.-J.},
  {\it Astrophys.~J.}, submitted (1997); preprint \hbox{astro-ph/9512121}.
\bibitem{lattboot:BoSaAh93} Boothroyd,~A.~I., Sackmann,~\hbox{I.-J.}, 
  \& Ahern,~S.~C., {\it Astrophys.~J.}, {\bf 416}, 762 (1993).
\bibitem{lattboot:BSW94} Boothroyd,~A.~I., Sackmann,~\hbox{I.-J.}, \&
  Wasserburg,~G.~J., {\it Astrophys.~J. (Lett.)}, {\bf 430}, L77 (1994).
\bibitem{lattboot:BSW95} Boothroyd,~A.~I., Sackmann,~\hbox{I.-J.}, \&
  Wasserburg,~G.~J., {\it Astrophys.~J. (Lett.)}, {\bf 442}, L21 (1995).
\bibitem{lattboot:Bre+93} Bressan,~A., Fagotto,~F., Bertelli,~G., \&
  Chiosi,~C., {\it Astron.\ Astrophys.\ Suppl.}, {\bf 100}, 647 (1993).
\bibitem{lattboot:Cetal71a} Castellani,~V., {\it et al.\/}, 
  {\it Astrophys.\ Space Sci.}, {\bf 10}, 340 (1971).
\bibitem{lattboot:Cetal71b} Castellani,~V., {\it et al.\/}, 
  {\it Astrophys.\ Space Sci.}, {\bf 10}, 355 (1971).
\bibitem{lattboot:CavSB96} Cavallo,~R.~M., Sweigart,~A.~V., \& Bell,~R.~A.,
  {\it Astrophys.~J. (Lett.)}, {\bf 464}, L79 (1996).
\bibitem{lattboot:CF88} Caughlan,~G.~R., \& Fowler,~W.~A.,
  {\it Atomic Data Nucl.\ Data Tables}, {\bf 40}, 205 (1988).
\bibitem{lattboot:Char94} Charbonnel,~C.,
  {\it Astron.\ Astrophys.}, {\bf 282}, 811 (1994).
\bibitem{lattboot:Char95} Charbonnel,~C.,
  {\it Astrophys.~J. (Lett.)}, {\bf 453}, L41 (1995).
\bibitem{lattboot:Dea92} Dearborn,~D.~S.~P.,
  {\it Phys.\ Reports}, {\bf 210}, 367 (1992).
\bibitem{lattboot:D84} Deupree,~R.~G., 1984, {\it Astrophys.~J.}, {\bf 287}, 
  268 (1984).
\bibitem{lattboot:DenW96} Denissenkov,~P.~A., \& Weiss,~A.,
  {\it Astron.\ Astrophys.}, {\bf 308}, 773 (1995).
\bibitem{lattboot:ElE94} El~Eid,~M.~F.,
  {\it Astron.\ Astrophys.}, {\bf 285}, 915 (1994).
\bibitem{lattboot:ForC97} Forestini,~M., \& Charbonnel,~C.,
  {\it Astron.\ Astrophys.\ Suppl.}, in press (1997).
\bibitem{lattboot:For+92} Forestini,~M., Goriely,~S., Jorissen,~A., \&
  Arnould,~M., {\it Astron.\ Astrophys.}, {\bf 261}, 157 (1992).
\bibitem{lattboot:Fetal91} Forestini,~M., Paulus,~G.,
  \& Arnould,~M., {\it Astron.\ Astrophys.}, {\bf 252}, 597 (1991).
\bibitem{lattboot:FreyLS96} Freytag,~B., Ludwig,~\hbox{H.-G.}, \&
  Steffen,~M., {\it Astron.\ Astrophys.}, {\bf 313}, 497 (1996).
\bibitem{lattboot:FL96a} Frost,~C.~A., \& Lattanzio,~J.~C., in
  {\it Stellar Evolution: What Should Be Done?}, Proc.\ of 32nd 
  Li\`ege Colloquium, ed.\ A.~Noels, D.~\hbox{Frepont-Caro}, M.~Gabriel,
  N.~Grevesse, \& P.~Demarque (U.~de Li\`ege: Belgium), p.~307 (1996).
\bibitem{lattboot:FL96b} Frost,~C.~A., \& Lattanzio,~J.~C.,
  {\it Astrophys.~J.}, {\bf 473}, 383 (1996).
\bibitem{lattboot:GBhere} Gallino,~R., \& Busso,~M., this volume.
\bibitem{lattboot:Gil89} Gilroy,~K.~K., {\it Astrophys.~J.},
  {\bf 347}, 835 (1989).
\bibitem{lattboot:GilB91} Gilroy,~K.~K., \& Brown,~J.~A.,
  {\it Astrophys.~J.}, {\bf 371}, 578 (1991).
\bibitem{lattboot:GdJ93} Groenewegen,~M.~A.~T., \& de~Jong,~T., 
  {\it Astron.\ Astrophys.}, {\bf 267} 410 (1993).
\bibitem{lattboot:HarL84a} Harris,~M.~J., \& Lambert,~D.~L.,
  {\it Astrophys.~J.}, {\bf 281}, 739 (1984).
\bibitem{lattboot:HarL84b} Harris,~M.~J., \& Lambert,~D.~L.,
  {\it Astrophys.~J.}, {\bf 285}, 674 (1984).
\bibitem{lattboot:Har+87} Harris,~M.~J., Lambert,~D.~L., Hinkle,~K.~H.,
  Gustafsson,~B., \& Eriksson,~K., {\it Astrophys.~J.}, {\bf 316}, 294 (1987).
\bibitem{lattboot:HarLS85} Harris,~M.~J., Lambert,~D.~L., \& Smith,~V.~V.,
  {\it Astrophys.~J.}, {\bf 299}, 375 (1985).
\bibitem{lattboot:HarLS88} Harris,~M.~J., Lambert,~D.~L., \& Smith,~V.~V.,
  {\it Astrophys.~J.}, {\bf 325}, 768 (1988).
\bibitem{lattboot:Herw+97} Herwig,~E., Bl\"ocker,~T., Sch\"onberner,~D., \&
  El~Eid,~M., {\it Astron.\ Astrophys.\ (Lett.)}, in press (1997).
\bibitem{lattboot:Hurl+94} Hurlburt,~N.~E., Toomre,~J., Massaguer,~J.~M., \&
  Zahn,~\hbox{J.-P.}, {\it Astrophys.~J.}, {\bf 421}, 245, (1994).
\bibitem{lattboot:Hus+94} Huss,~G.~R., Fahey,~A.~J., Gallino,~R., \&
  Wasserburg,~G.~J., {\it Astrophys.~J. (Lett.)}, {\bf 430}, L81 (1994).
\bibitem{lattboot:Hus+92} Huss,~G.~R., Hutcheon,~I.~D., Wasserburg,~G.~J., \&
  Stone,~J., {\it Lunar Planet.\ Sci.}, {\bf 23}, 563 (1992).
\bibitem{lattboot:Iben83} Iben, I.~Jr., {\it Astrophys.~J. (Lett.)},
  {\bf 275}, L65 (1983).
\bibitem{lattboot:IR82a} Iben, I.~Jr., \& Renzini, A.,
  {\it Astrophys.~J. (Lett.)}, {\bf 259}, L79 (1982).
\bibitem{lattboot:IR82b} Iben, I.~Jr., \& Renzini, A.,
  {\it Astrophys.~J. (Lett.)}, {\bf 263}, L231 (1982).
\bibitem{lattboot:IR83} Iben, I.~Jr., \& Renzini, A.,
  {\it Ann.\ Rev.\ Astron.\ Astrophys.}, {\bf 21}, 271 (1983). 
\bibitem{lattboot:Kah92} Kahane,~C, Cernicharo,~J.,
  \hbox{G\'omez-Gonz\'alez},~J., \& Gu\'elin,~M., {\it Astron.\ Astrophys.},
  {\bf 256}, 235 (1992).
\bibitem{lattboot:Kra94} Kraft,~R.~P., {\it Proc.\ Astron.\ Soc.\ Pacific},
  {\bf 106}, 553 (1994).
\bibitem{lattboot:Kra+93} Kraft,~R.~P., Sneden,~C., Langer,~G.~E.,
  \& Shetrone,~M.~D., {\it Astron.~J.}, {\bf 104}, 645 (1993).
\bibitem{lattboot:Kudr78} Kudritzki,~R.~P.,~\& Reimers,~D.,
  {\it Astron.\ Astrophys.}, {\bf 70}, 227 (1978).
\bibitem{lattboot:La90} Landr\'e,~V., Prantzos,~N., Aguer,~P., Bogaert,~G.,
  Lefebvre,~A., \& Thibaud,~J.~P., {\it Astron.\ Astrophys.},
  {\bf 240}, 85 (1990)
\bibitem{lattboot:LanH95} Langer,~G.~E., \& Hoffman,~R.~D.,
  {\it Proc.\ Astron.\ Soc.\ Pacific}, {\bf 107}, 1177 (1995).
\bibitem{lattboot:LanHS93} Langer,~G.~E., Hoffman,~R., \& Sneden,~C.,
  {\it Proc.\ Astron.\ Soc.\ Pacific}, {\bf 105}, 301 (1993).
\bibitem{lattboot:Lang+86} Langer,~G.~E., Kraft,~R.~P., Carbon,~D.~F., \&
  Friel,~E., {\it Proc.\ Astron.\ Soc.\ Pacific}, {\bf 98}, 473 (1986).
\bibitem{lattboot:L92} Lattanzio,~J.~C., {\it Proc.\ Astron.\ Soc.\ Aust.}, 
  {\bf 10}, 120L (1992).
\bibitem{lattboot:Letal96} Lattanzio,~J.~C., Frost,~C.~A., Cannon,~R.~C., 
  \& Wood,~P.~R., {\it Mem.\ Soc.\ Astron.\ Italia.}, {\bf 67}, 729 (1996).
\bibitem{lattboot:Letal97} Lattanzio,~J.~C., Frost,~C.~A., Cannon,~R.~C., 
  \& Wood,~P.~R., in preparation (1997).
\bibitem{lattboot:L71} Lauterborn,~D., {\it et al.\/},
  {\it Astron.\ Astrophys.}, {\bf 10}, 97 (1971).
\bibitem{lattboot:MarBC96} Marigo,~P., Bressan,~A., \& Chiosi,~C.,
  {\it Astron.\ Astrophys.}, {\bf 313}, 545 (1996).
\bibitem{lattboot:Metal96} Mowlavi,~N., Jorissen,~A., \& Arnould,~M.,
  {\it Astron.\ Astrophys.}, {\bf 311}, 803 (1996).
\bibitem{lattboot:Nit97} Nittler,~L.~R., preprint (1997).
\bibitem{lattboot:Nit+94} Nittler,~L.~R., Alexander,~C.~M.~O'D., Gao,~X.,
  Walker,~R.~M., \& Zinner,~E.~K., {\it Nature}, {\bf 370}, 443 (1994).
\bibitem{lattboot:Pin+89} Pinsonneault,~M.~H., Kawaler,~S.~D., Sofia,~S., \&
  Demarque,~P., {\it Astrophys.~J.}, {\bf 338}, 424 (1989).
\bibitem{lattboot:Rei75} Reimers,~D., in {\it Problems in Stellar Atmospheres
  and Envelopes}, ed.\ B.~Bascheck, W.~H.~Kegel, \& G.~Traving (New York:
  Springer), 229 (1975).
\bibitem{lattboot:RenV81} Renzini,~A., \& Voli,~M., {\it Astron.\ Astrophys.},
  {\bf 94}, 175 (1981).
\bibitem{lattboot:RieuZ95} Rieutord,~M., \& Zahn,~\hbox{J.-P.},
  {\it Astron.\ Astrophys.}, {\bf 296}, 127 (1995).
\bibitem{lattboot:SB91} Sackmann,~\hbox{I.-J.}, \& Boothroyd,~A.~I.,
  {\it Astrophys.~J.}, {\bf 366}, 529 (1991).
\bibitem{lattboot:SackBoot92} Sackmann,~\hbox{I.-J.}, \& Boothroyd,~A.~I.,
  {\it Astrophys.~J. (Lett.)}, {\bf 392}, L71 (1992).
\bibitem{lattboot:SB97} Sackmann,~\hbox{I.-J.}, \& Boothroyd,~A.~I.,
  {\it Astrophys.~J.}, submitted (1997); preprint \hbox{astro-ph/9512122}.
\bibitem{lattboot:SBF90} Sackmann,~\hbox{I.-J.}, Boothroyd,~A.~I., \&
  Fowler,~W.~A., {\it Astrophys.~J.}, {\bf 360}, 727 (1990).
\bibitem{lattboot:SBK93} Sackmann,~\hbox{I.-J.}, Boothroyd,~A.~I., \&
  Kraemer,~K.~E., {\it Astrophys.~J.}, {\bf 418}, 457 (1993).
\bibitem{lattboot:SSD74} Sackmann,~\hbox{I.-J.}, Smith,~R.~L., \&
  Despain,~K.~H. {\it Astrophys.~J.}, {\bf 187}, 555 (1974).
\bibitem{lattboot:SDU75} Scalo,~J.~M., Despain,~K.~H., \& Ulrich,~R.~K., 
  {\it Astrophys.~J.}, {\bf 196} 805 (1975).
\bibitem{lattboot:Schal+92} Schaller,~G., Schaerer,~D., Meynet,~G., \&
  Maeder,~A., {\it Astron.\ Astrophys.\ Suppl.}, {\bf 96}, 269 (1992).
\bibitem{lattboot:ST92} Smith,~G.~H., \& Tout,~C.~A., 
  {\it Mon.\ Not.\ Royal Astron.\ Soc.}, {\bf 256} 449 (1992).
\bibitem{lattboot:SL89} Smith,~V.~V., \& Lambert,~D.~L., {\it Astrophys.~J.},
  {\bf 345}, 375 (1989).
\bibitem{lattboot:SL90} Smith,~V.~V., \& Lambert,~D.~L.,
  {\it Astrophys.~J. (Lett.)}, {\bf 361}, L69 (1990).
\bibitem{lattboot:Smithetal95} Smith,~V.~V., Lambert,~D.~L., Plez,~B., \&
  Lubowich,~D.~A., {\it Astrophys.~J.}, {\bf 441}, 735 (1995).
\bibitem{lattboot:Stran+97} Straniero,~O., Chieffi,~A., Limongi,~M.,
  Busso.,~M., Gallino,~R., \& Arlandini,~C., {\it Astrophys.~J.},
  {\bf 478}, 332 (1997).
\bibitem{lattboot:Setal95} Straniero,~O., Gallino,~R., Busso.,~M.,
  Chieffi,~A., Limongi,~M., \& Salaris,~M., {\it Astrophys.~J. (Lett.)},
  {\bf 440}, L85 (1995).
\bibitem{lattboot:SweM79} Sweigart,~A.~V., \& Mengel,~J.~G.,
  {\it Astrophys.~J.}, {\bf 229}, 624 (1979).
\bibitem{lattboot:TalZ97} Talon,~S., \& Zahn,~\hbox{J.-P.}, {\it Astron.\
  Astrophys.}, {\bf 317}, 749 (1997).
\bibitem{lattboot:Tim95} Timmes,~F.~X., private communication (1995).
\bibitem{lattboot:TimWW95} Timmes,~F.~X., Woosley,~S.~E., \& Weaver,~T.~A.,
  {\it Astrophys.~J. Suppl.}, {\bf 98}, 617 (1995).
\bibitem{lattboot:vdHG97} van~den~Hoek,~L.~B., \& Groenewegen,~M.~A.~T.,
  {\it Astron.\ Astrophys.\ Suppl.}, in press; preprint
  \hbox{astro-ph/9610030} (1997).
\bibitem{lattboot:VW93} Vassiliadis,~E. \& Wood,~P.~R., {\it Astrophys. J.}, 
  {\bf 413}, 641 (1993).
\bibitem{lattboot:WBS95} Wasserburg,~G.~J., Boothroyd,~A.~I., \&
  Sackmann,~\hbox{I.-J.}, {\it Astrophys.~J. (Lett.)}, {\bf 447}, L37 (1995).
\bibitem{lattboot:Was+94} Wasserburg,~G.~J., Busso,~M., Gallino,~R., \&
  Raiteri,~C.~M., {\it Astrophys.~J.}, {\bf 424}, 412 (1994).
\bibitem{lattboot:WeaW93} Weaver,~T.~A, \& Woosley,~S.~E.,
  {\it Phys.\ Rept.}, {\bf 227}, 65 (1993).
\bibitem{lattboot:W81} Wood,~P.~R., {\it Astrophys.~J.}, {\bf 248}, 311 (1981).
\bibitem{lattboot:Zahn92} Zahn,~\hbox{J.-P.}, {\it Astron.\ Astrophys.},
  {\bf 265}, 115 (1992).

\end{references}
\end{document}